%% file: pvis-topopilot-main.tex
\documentclass{vgtc}                   
\onlineid{3720}

\usepackage{times} 

\usepackage{enumitem}
\usepackage{algorithm}
\usepackage{algpseudocode}
\usepackage{amsmath}
\usepackage{amsfonts}
\usepackage{listings}
\usepackage{xcolor}
\usepackage{mathptmx}
\usepackage[most]{tcolorbox}

\newtcolorbox{highlight}{
    colback=white,
    colframe=black,
    boxrule=0.5pt,
    arc=0pt,
    width=\columnwidth,
    left=6pt,
    right=6pt,
    top=4pt,
    bottom=4pt
}

\usepackage{adjustbox}

\newcommand{\FlexibilityR}{\textbf{R1}}
\newcommand{\ExtensibilityR}{\textbf{R2}}
\newcommand{\ReproducibilityR}{\textbf{R3}}
\newcommand{\InterpretabilityR}{\textbf{R4}}

\newcommand{\ClarificationFailureF}{\textbf{F1}}
\newcommand{\ConfusionOfCapabilitiesF}{\textbf{F2}}
\newcommand{\BadParameterValuesF}{\textbf{F3}}
\newcommand{\InvalidWorkflowF}{\textbf{F4}}
\newcommand{\IntentionsNotSatisfiedF}{\textbf{F5}}
\newcommand{\ErroneousExecutionF}{\textbf{F6}}

\newcommand{\para}[1]{\noindent{\textbf{#1}}}
\newcommand{\tool}{\emph{TopoPilot}}
\newcommand{\R}{\mathbb{R}}
\newcommand{\X}{\mathbb{X}}
\newcommand{\T}{\mathbb{T}}

\newcommand{\myuser}[1]{\textcolor{ForestGreen}{\textit{#1}}}
\newcommand{\myagent}[1]{\textcolor{Blue}{\textit{#1}}}
\newcommand{\usercolor}[1]{\textcolor{ForestGreen}{#1}}
\newcommand{\agentcolor}[2]{\textcolor{Blue}{#1}}

\definecolor{bulgarianrose}{rgb}{0.28, 0.02, 0.03}
\newcommand{\prompt}[1]{\textcolor{bulgarianrose}{#1}}

\newcommand{\newcolor}{black}
\newcommand{\new}[1]{\textcolor{\newcolor}{#1}}

\newcommand{\inlineicon}[1]{%
  \raisebox{-0.2em}{\includegraphics[height=1em]{#1}}%
}

\graphicspath{{./}} 

\vgtccategory{Research}
\vgtcinsertpkg

\title{\textbf{TopoPilot}: Reliable Conversational Workflow Automation for Topological Data Analysis and Visualization}

\author{Nathaniel Gorski \thanks{e-mail: gorski@sci.utah.edu}\\ \scriptsize University of Utah
\and Shusen Liu\thanks{e-mail: liu42@llnl.gov} \\ \scriptsize Lawrence Livermore National Laboratory
\and Bei Wang\thanks{e-mail:beiwang@sci.utah.edu} \\ \scriptsize University of Utah
}

\teaser{
\centering
\includegraphics[width=1.0\linewidth]{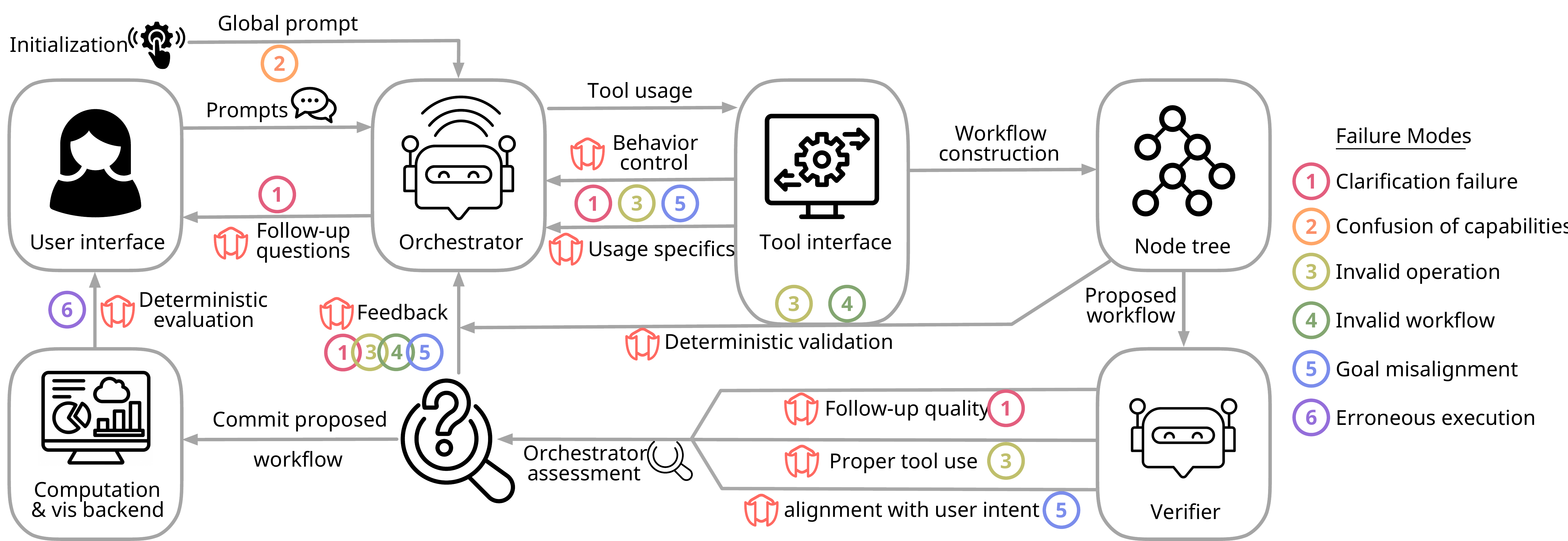}
\vspace{-6mm}
\caption{System overview of {\tool}, illustrating its end-to-end conversational workflow and evaluation loop. Elements marked with the safeguard icon \inlineicon{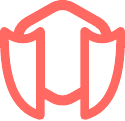} denote mechanisms for ensuring reliability, mitigating six possible failure modes \inlineicon{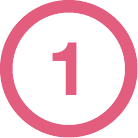} through  \inlineicon{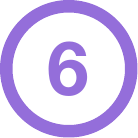}. Users initiate interaction through a conversational interface, engaging the orchestrator agent, which invokes tools via a tool interface (such as MCP) to construct a workflow in a \emph{node tree}---a custom data structure for multi-step visualization pipelines with built-in deterministic safeguards. The resulting workflow is validated by a verifier agent to ensure proper tool usage, completeness of information, and alignment with user intent. Upon successful verification, the workflow is deterministically executed and rendered in the user interface.}
\label{fig:teaser}
}

\abstract{
\input{sec-abstract.tex}
}


\nocopyrightspace

\begin{document}

\maketitle

\input{sec-introduction.tex}
\input{sec-related-work.tex}
\input{sec-requirements.tex}
\input{sec-overview.tex}
\input{sec-method.tex}

\input{sec-case-studies.tex}
\input{sec-results.tex}
\input{sec-expert.tex}
\input{sec-limitations.tex}

\input{sec-ack.tex}

\bibliographystyle{abbrv-doi-hyperref-narrow}
\bibliography{refs-topopilot}

\newpage
\appendix

\crefalias{section}{appendix}
\crefalias{subsection}{appendix}

\renewcommand{\myuser}[1]{\noindent\textcolor{ForestGreen}{\textit{#1}}}
\renewcommand{\myagent}[1]{\noindent\textcolor{Blue}{\textit{#1}}}

\input{appendix-dataset-descriptions.tex}
\input{appendix-descriptors.tex}

\input{appendix-additional-results.tex}
\input{appendix-additional-case-studies.tex}
\input{appendix-implementation-details.tex}
\input{appendix-tester.tex}
\input{appendix-evaluation-notes.tex}

\input{appendix-correctness.tex}
\input{appendix-chat-logs.tex}
\input{appendix-evaluation-prompts.tex}

\end{document}

%% file: sec-abstract.tex
Recent agentic systems demonstrate that large language models can generate complex codebases from natural language prompts. Despite this progress, they present reliability concerns for scientific visualization due to invalid operations, subtle errors, and failure to resolve underspecified inputs. In this work, we introduce {\tool}, a reliable and extensible agentic framework for automating complex scientific visualization workflows. We focus on topological data analysis and visualization as a primary use case, while designing the framework to generalize to broader visualization domains.\\
\indent {\tool} employs a two-agent architecture: an orchestrator translates user prompts into workflows composed of atomic backend actions, while a verifier evaluates workflows prior to execution. Its modular design isolates components and supports new topological descriptors and domain-specific workflows without modifying the core system. It also employs numerous deterministic guardrails to ensure semantic consistency and accurate execution. To systematically improve reliability, we develop a taxonomy of failure modes and employ targeted safeguards for each class of error. Across 2{,}000 simulated multi-turn conversations spanning 100 prompts, including adversarial and infeasible requests, {\tool} achieves a success rate exceeding 95\%, compared with under 50\% for a baseline without comprehensive guardrails and verification. We also show that {\tool} outperforms Claude Code and Codex on similar prompts in terms of accuracy, completion time, and token usage.

%% file: sec-introduction.tex
\section{Introduction}
\label{sec:introduction}

Coding agents such as Codex and Claude Code enable users to translate natural-language instructions into complex codebases with limited manual intervention. Scientific visualization is a promising application for these agents because its workflows often require substantial programming and visual expertise, frequently involving data preprocessing and domain-specific analysis. Automating these workflows could therefore reduce the effort and expertise required to produce scientific visualizations and make advanced visualization techniques more accessible to novice users.

However, coding agents often perform poorly on scientific visualization tasks even when they succeed on conventional programming problems~\cite{ai2026scivisagentbench}. Agents may generate executable code that misapplies visualization techniques or produces incorrect or misleading results, particularly when underspecified prompts require assumptions about the intended analysis or representation. Such errors are difficult to detect by inspecting visualizations alone, and require analyzing the dense scripts used to generate them. This process is cumbersome even for expert users. Moreover, repeated agent interaction and correction can be computationally expensive, with complex visualization tasks potentially consuming millions of tokens.

Recent agentic visualization systems such as VizGenie~\cite{biswas2025vizgenie}, ChatVis~\cite{mallick2024chatvis,peterka2025chatvis}, and ParaView-MCP~\cite{liu2025paraview} reduce some of these risks by providing domain-specific agentic harnesses. Systems such as VizGenie and ChatVis nevertheless rely on code generation to construct visualizations and therefore remain susceptible to similar reliability concerns as coding agents. Tool-based systems such as ParaView-MCP instead constrain agents to predefined atomic operations through the Model Context Protocol (MCP), eliminating many failure modes associated with unconstrained code generation. However, constraining the available operations does not guarantee that an agent will compose them into a correct workflow. It may still select inappropriate operations, arrange otherwise valid operations in the wrong sequence, or choose parameters that are inconsistent with the user’s analytical intent. Moreover, existing agentic visualization systems have largely focused on relatively simple workflows, such as surface or volume rendering. Although useful, these tasks capture only a fraction of the complexity of real-world scientific visualization, where workflows often involve many interdependent stages and errors introduced at one stage can propagate through subsequent analysis and rendering steps.

Topological data analysis and visualization exemplifies this challenge. Topological methods have become powerful tools for analyzing scientific data, enabling researchers to uncover subtle structural patterns across diverse domains (e.g., ~\cite{bilsky2025understanding, servis2022amphiphile, hu2021persistent}). Unlike the relatively simple visualization tasks considered by many existing agentic systems, topological workflows are often inherently multi-stage, involving preprocessing, computation and simplification of topological descriptors, and often downstream analysis such as temporal tracking with optimal transport. Reliably automating such workflows not only requires selecting valid individual operations, but also correctly composing and parameterizing those operations across the entire analysis pipeline.

In this work, we introduce {\tool}, a reliable and extensible agentic framework for automating complex scientific visualization workflows. {\tool} provides a scaffolded agentic system for constructing workflows, including creating and correctly sequencing the stages of a visualization pipeline. Each stage is executed using pre-written code to maximize reliability. We use topological data analysis and visualization as a representative use case while designing the framework to generalize to a broader range of visualization domains.

As illustrated in \cref{fig:teaser}, {\tool} adopts a reliability-centered two-agent architecture. The orchestrator agent translates user prompts into workflows composed of atomic actions defined in a custom backend (e.g.,~via MCP), while the verifier agent validates these workflows before execution to ensure that they align with the users' intentions. {\tool} employs a custom workflow representation, called the \emph{node tree}, that enforces workflow-level semantic correctness while preserving flexibility and enabling the seamless integration of diverse libraries within a unified execution model.

To further enhance reliability, {\tool} incorporates feature-specific deterministic guardrails that constrain the orchestrator's behavior for particular node types and mitigate a range of failure modes. These guardrails  resolve common linguistic ambiguities, enforce reasonable default design choices, and ensure the user provides all necessary information. Finalized workflows can be exported as portable Python command-line tools that use predefined macros, enabling users to easily verify workflow correctness.

We present extensive empirical evaluations demonstrating that {\tool} reliably achieves its intended outcomes. We also provide a comparison with existing coding agents showing that {\tool} achieves superior accuracy, running time, and token usage.
Our contributions include:
\begin{itemize}[noitemsep,leftmargin=*]
\item \textbf{A reliability-focused agentic system for topology-based visualization.} We introduce {\tool}, a conversational agentic framework that accurately automates end-to-end topology-based visualization workflows through natural language interaction.

\item \textbf{A flexible and extensible computation engine with deterministic evaluation.} {\tool} employs a custom backend that integrates atomic operations organized through a novel \emph{node tree} representation. This architecture ensures that any pipeline ordering is semantically valid and each stage executes correctly, allowing for complex workflows to be produced using diverse libraries and frameworks.

\item \textbf{A taxonomy of failure modes with targeted mitigation strategies.} We systematically enumerate potential failure modes within the {\tool} framework and implement explicit safeguards---guardrails for agents and complementary mitigation strategies---to address each class of error, strengthening reliability across diverse usage scenarios.

\item \textbf{A large-scale empirical evaluation of system reliability.} We conduct 2,000 simulated multi-turn conversations across 100 distinct prompts, demonstrating {\tool}'s reliable performance across a broad range of capabilities. We further conduct 384 trials using Claude Code and Codex on the same prompts, highlighting the reliability advantages of {\tool}'s purpose-built framework.

\end{itemize}

%% file: sec-related-work.tex
\section{Related Work}
\label{sec:related-work}

\subsection{Natural Language Visualization Tools}
The use of AI in visualization has gained significant momentum in recent years; see \cite{wu2021ai4vis} for an early survey. LLMs such as ChatGPT and Claude are now widely used to generate visualization code. However, this code-generation approach is often error-prone, frequently yielding undesirable visual encodings or even code that fails to run. As a result, custom AI-powered systems have emerged to produce more reliable, curated visualizations. These systems typically employ AI agents, semi-autonomous components constructed by wrapping an LLM with targeted prompts, and often coordinate multiple agents with specialized roles within a multi-agent framework. Many also integrate vision models to inspect generated visualizations and iteratively refine them based on visual feedback.

A number of AI tools have been developed specifically for scientific visualization. T2TF \cite{jeong2024text} offers a natural-language interface for designing transfer functions using differentiable optimization, while FlowNL \cite{huang2022flownl} introduces a custom flow-visualization grammar and a translator that maps natural-language to this syntax. Other systems rely on AI agents to tune visualization parameters: AVA \cite{liu2024ava} assigns separate agents to distinct visualization types and refines results using both prompts and visual feedback; NLI4VolVis \cite{ai2025nli4volvis} employs a multi-agent design for volume visualization and integrates Novel View Synthesis (NVS) for efficient rendering. Although effective for their specialized tasks, these systems focus primarily on direct visualization of scalar fields and do not generalize easily to more complex or multi-step analytical workflows.

To enhance flexibility, several tools instead generate visualization code. ChatVis \cite{mallick2024chatvis, peterka2025chatvis} uses a multi-agent architecture with Retrieval-Augmented Generation (RAG) to produce ParaView scripts, iteratively refining code using error messages. VizGenie \cite{biswas2025vizgenie} follows a similar model but adds fine-tuned vision systems and labeled visual data to allow natural-language adjustment of visualization attributes such as transfer functions. While highly flexible, such systems are resource-intensive: repeated code generation and error correction consume many tokens, and subtle implementation mistakes can arise even in syntactically correct code. These issues are compounded when prompts are ambiguous, as LLMs tend to infer missing details rather than request clarification.

Among existing systems, {\tool} is most closely related to ParaView-MCP~\cite{liu2025paraview}, which similarly avoids code generation by exposing selected ParaView GUI capabilities through an agentic natural-language interface. Although ParaView-MCP offers broad visualization functionality, it is constrained by a fixed and limited toolset, is difficult to extend, and lacks systematic validation mechanisms, allowing user errors or invalid configurations to propagate unchecked. In contrast, {\tool} is built on a flexible and extensible backend that emphasizes modularity and incorporates rigorous guardrails, making it better suited for complex, multi-stage, and evolving analytical workflows.

Recently, SciVisAgentBench \cite{ai2026scivisagentbench} was introduced to benchmark scientific visualization agents on a selected set of tasks. Its evaluation showed that ChatVis and ParaView-MCP generally used substantially fewer tokens than Claude Code, although Claude Code achieved higher accuracy. It also evaluated an earlier version of {\tool} on a task set that included many capabilities outside its intended scope, \new{such as working with unsupported data formats}.
In contrast, our evaluation focuses on complex, end-to-end workflows specifically aligned with {\tool}'s capabilities and extends beyond the workflows considered in SciVisAgentBench.

Multi-agent architectures have also been explored in information visualization, including VisPath \cite{seo2025automated}, PlotGen \cite{goswami2025plotgen}, MatPlotAgent \cite{yang2024matplotagent}, and CoDA \cite{chen2025coda}. Other systems, such as YAC~\cite{lange2025generative} and the framework of Gyarmati et al.~\cite{ferenc2025composable}, provide custom backends through which agents perform visualization tasks. Related agentic approaches have also been applied to visual analytics \cite{zhao2025proactiveva} and visual storytelling \cite{islam2024datanarrative}.

\subsection{Reliability of Agentic Frameworks}

A substantial body of work has investigated methods for improving the reliability of LLMs, including techniques to reduce hallucinations and enhance reasoning (see~\cite{ayyamperumal2024current} for a survey). In contrast, comparatively less attention has been devoted to the reliability of agentic systems, where errors may compound across multi-step workflows and tool invocations. Several studies have proposed mechanisms to mitigate failure when coordinating teams of agents~\cite{patel2026six, xian2025reliable}. In contrast, {\tool} adopts a deliberately simple two-agent architecture to reduce coordination complexity. Other approaches dynamically represent tasks as trees or graphs, where nodes correspond to atomic operations—an idea conceptually similar to our \emph{node tree} representation—and have shown that such structured decompositions can improve reliability~\cite{patel2026six, balis2026research}. However, \new{in \cite{patel2026six}}, atomic operations are themselves delegated to additional agents or agent teams, whereas {\tool} executes them deterministically to enforce correctness guarantees. \new{Introduced in our original preprint, {\tool} represents reproducible workflows as (directed acyclic) graphs and executes them deterministically; subsequent work \cite{balis2026research} uses a similar approach.} Additional strategies aim to improve the accuracy of tool use by LLMs~\cite{liu2025toolplanner, liu2026toolgate}, improve the quality of agentic planning \cite{wang2026planning}, or provide better feedback to agentic systems \cite{liu2026scifi, zou2026amtfv}.



%% file: sec-requirements.tex
\section{System Requirements and Failure Modes}
\label{sec:requirements}

\new{We aim to design a reliable and extensible agentic system that automates complex workflows for domain scientists. To make {\tool} useful in real-world scientific settings, we identify several essential design requirements. Scientific pipelines often require repeated iterations, so the system should support incremental workflow refinement while avoiding costly recomputation. Workflows also incorporate diverse tools, so the system should readily accommodate new tools and frameworks. These considerations motivate two design requirements:}

\para{\FlexibilityR. Flexibility.} The system should allow users to adjust or refine visualizations with minimal recomputation whenever possible.

\para{\ExtensibilityR. Extensibility.} The system should easily incorporate diverse libraries and algorithms, with new features simple to integrate without modifying the core architecture.

\new{In practice, scientific visualization scripts are often reused across datasets. Supporting this reuse motivates the following design requirement:}

\para{\ReproducibilityR. \new{Reusability.}} Visualization workflows should be exportable to external tools so workflows can be reapplied to similar datasets.

\new{Finally, errors in scientific visualization pipelines can produce plausible but incorrect visualizations, making them difficult to detect from the final output alone. Users therefore need a clear way to inspect and verify the computations performed by the model. This motivates our final design requirement:}


\para{\InterpretabilityR. Interpretability.} The system should expose the operations, parameters, and execution steps used to produce a visualization in a form that users can inspect and evaluate.

\new{Beyond these requirements, {\tool} must operate reliably. Agentic systems can fail in many ways, as documented in prior work (e.g., \cite{mallick2024chatvis}). We address six failure modes that broadly affect visualization systems, some of which are especially problematic for topology-based systems. Agentic visualization systems typically generate visualizations by executing workflows developed through dialogue with users. Failures can arise during this dialogue, workflow creation, or workflow execution. We classify these failure modes as chat failures, workflow failures, or execution failures, and describe them below. We describe targeted safeguards in \cref{sec:method-failure-modes}}.

\para{\new{Chat failures:}} \new{Before constructing a workflow, the system must work with the user to establish a feasible, sufficiently specified objective. This requires accurately communicating its capabilities and resolving ambiguities in the user's request. These concerns are especially important in topological settings, where subtle differences in wording can imply different, potentially costly computations. These challenges motivate two failure modes below:}

\para{\ClarificationFailureF. Clarification failure.} The system fails to request or obtain essential information from the user, such as if the prompt is underspecified or ambiguous.

\para{\ConfusionOfCapabilitiesF. Confusion of capabilities.} The system \new{suggests a task beyond its supported functionality, or incorrectly reports that a feasible task lies outside its capabilities.}

\para{\new{Workflow failures:}} \new{Once the user has specified their request, the model constructs a workflow. This process can fail in three ways. First, the model may select invalid or suboptimal operations, for example, by choosing inappropriate parameter values. Second, it may combine valid operations in an invalid sequence. Third, it may construct a valid workflow that does not fulfill the user's intent. These scenarios define three failure modes:}

\para{\BadParameterValuesF. Invalid \new{operation.}} The agent selects an unsupported algorithm or supplies a parameter value that is illegal or produces very poor results.

\para{\InvalidWorkflowF. Invalid workflow.} The system constructs illegal sequences of operations.

\para{\IntentionsNotSatisfiedF. Goal misalignment.} \new{The system executes a workflow that does not align with a valid interpretation of the user's request.}

\para{\new{Execution failure:}} \new{Finally, a proper workflow may have implementation errors, leading to an incorrect result. This scenario motivates our final failure mode:}

\para{\ErroneousExecutionF. Erroneous execution.} \new{A valid workflow fails during execution due to implementation errors}, resulting in incorrect outputs, runtime failures, or corrupted artifacts.

%% file: sec-overview.tex
\section{System Overview}
\label{sec:overview}

In this section, we provide an overview of the {\tool} framework. We describe the system architecture and explain how it satisfies the design requirements in \cref{sec:overview-design-requirements}, and we summarize its key features in \cref{sec:overview-features}. Detailed implementation specifics are presented in \cref{sec:method} and in the supplement.

\subsection{Satisfaction of Design Requirements}
\label{sec:overview-design-requirements}

Built using LLM tools, {\tool} can be paired with any compatible LLM (e.g., ChatGPT or Claude) and is distributed as a Python library that runs in environments with ParaView and the Topology Toolkit~\cite{tierny2018ttk}. The system is designed to emulate the workflow of a topology expert advising a domain scientist. In this interaction, the user may begin with foundational questions and exploratory visualizations, while the advisor asks clarifying questions, provides contextual guidance, and refines the analysis. For computationally intensive tasks, the advisor may first operate on data subsets before scaling to the full dataset. Once the user is satisfied with the exploratory results, the advisor helps produce a finalized script applicable to multiple datasets. {\tool} mirrors this expert-guided interaction while satisfying the requirements outlined in \cref{sec:requirements}.

\begin{figure}[!ht]
\includegraphics[width=\linewidth]{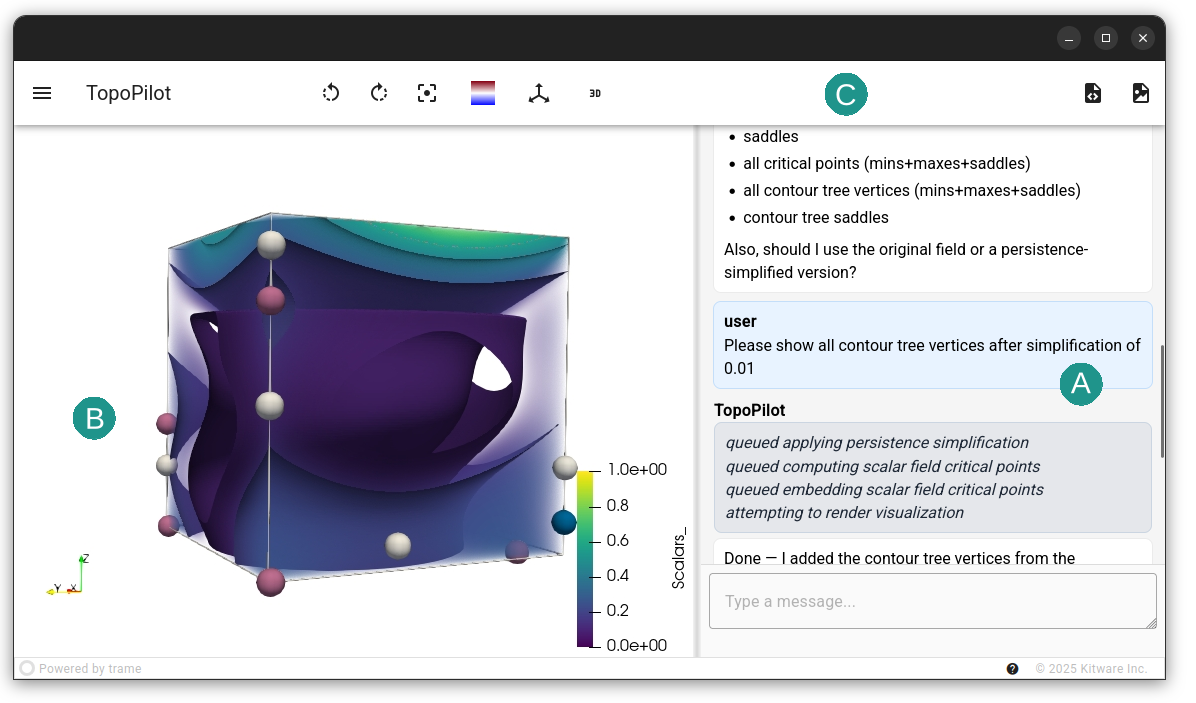}
\vspace{-6mm}
\caption{A screenshot of the visualization window used to interact with {\tool}. (A) a built-in chat interface is used to communicate with the LLM. (B) An interactive visualization pane displays the generated visualization. (C) A toolbar provides basic camera and visualization controls, as well as the ability to save screenshots, videos (for time-varying data), and auto-generated Python code.}
\vspace{-6mm}
\label{fig:panel}
\end{figure}

\para{Visual interface.} Users interact with {\tool} through a browser-based interface, shown in \cref{fig:panel}. Visualizations are generated by submitting natural language requests through an embedded chat window. The conversational interface is structured so that {\tool} requests any necessary follow-up information before constructing a workflow, ensuring that all required inputs are specified prior to execution. The resulting visualization is displayed in an interactive rendering pane that occupies the majority of the interface.

The interface also includes a toolbar providing essential visualization controls, such as camera manipulation and toggling of orientation axes and color bars. Additional options allow users to export images, videos (for time-varying data), or Python code. Exporting Python code produces a portable command-line script that reproduces the visualization in any environment where the {\tool} Python library is installed, satisfying {\ReproducibilityR}. The script invokes predefined macros to reconstruct the primary computation pipeline, and it explicitly exposes all parameter choices in a clear and concise format. This design ensures that the workflow is transparent and easily interpretable, satisfying {\InterpretabilityR}. An example of the exported code is shown in \cref{fig:code-export-example}.

\begin{figure}[h]
\centering
\vspace{-4mm}
\begin{lstlisting}[basicstyle=\fontsize{6.5}{7.5}\ttfamily,xleftmargin=0pt,xrightmargin=0pt]
scalar0 = LoadScalarField(load_path='Isabel.vti', 
                          arrayName='Scalars_')
scalar1 = PersistenceSimplification(parent_id = scalar0.id, 
                                    epsilon = 0.04,
                                    abs_threshold = False)
pd2 = ComputePersistenceDiagram(parent_id = scalar1.id)
pd_embed3 = EmbedPersistenceDiagram(parent_id = pd2.id, 
                                    ball_radius = 0.02, 
                                    tube_radius = 0.01)
\end{lstlisting}
\vspace{-4mm}
\caption{Python code generated to reconstruct a workflow for computing a simplified persistence diagram; some names are shortened due to space constraints.}
\label{fig:code-export-example}
\vspace{-4mm}
\end{figure}

\para{Custom backend.}~{\tool} operates as a two-agent architecture. The \textit{orchestrator} creates visualizations while the \textit{verifier} validates the orchestrator's actions. The orchestrator creates visualizations by interacting with a custom computation engine (described in detail in \cref{sec:method-implementation}) and constructs a visualization workflow by chaining together atomic operations. Deterministic safeguards ensure that the generated workflows are semantically valid and execute properly. Additional prompt-based methods and the verifier agent help resolve linguistic ambiguities and ensure the user's intention is satisfied. All intermediate computations are cached, allowing visualization parameters, such as color maps or glyph sizes, to be modified without recomputing upstream results, thereby satisfying {\FlexibilityR}.

Each operation is implemented as a standalone Python class that inherits from a shared abstract base class. New capabilities can therefore be introduced by defining additional subclasses, without modifying existing components. Operations consume input data and either transform it into new representations or produce visual outputs independent of prior steps. This modular design permits the integration of diverse libraries and frameworks, enabling new functionality to be incorporated seamlessly alongside existing operations and satisfying {\ExtensibilityR}.

\subsection{Various Features Supported by {\tool}}
\label{sec:overview-features}

\para{Diverse data types and visual encodings.}
{\tool} supports scalar, vector, and tensor field data, including 2D and 3D scalar fields and 2D vector and tensor fields. 2D scalar fields are visualized as heatmaps, while 3D scalar fields are rendered via volume rendering or isocontours. As volume rendering is not a primary focus, we omit iterative transfer-function refinement and instead provide a lightweight transfer-function selector (see the supplement for details). For 2D vector fields, {\tool} supports LIC visualization, and for 2D symmetric tensor fields, a variation of hyperLIC~\cite{zheng2003hyperlic}. Asymmetric tensor fields are visualized using eigenvector or eigenvalue partitions. Colormap can be specified via natural language.

\begin{figure}[!ht]
\centering
\vspace{-3mm}
\includegraphics[width=0.8\linewidth]{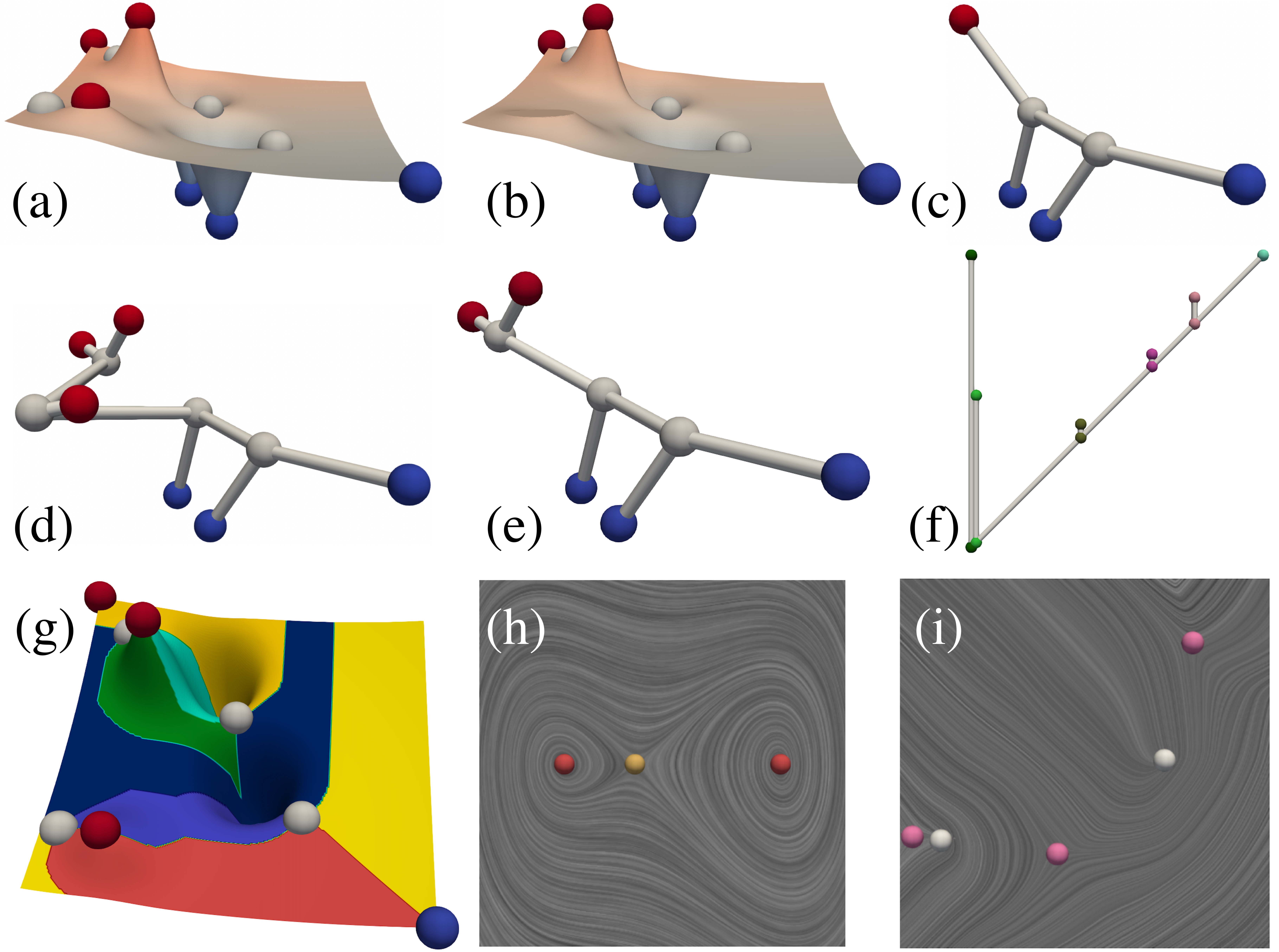}
\vspace{-2mm}
\caption{Examples of topological descriptors supported by {\tool}.
(a) The graph of a 2D function $f$ with its critical points: maxima in red, saddles in white, and minima in blue. (b) $f$ with a saddle-maximum pair removed after persistence simplification. 
(c) Merge tree of $f$.
(d)–(e) Contour tree of $f$ before and after persistence simplification. 
(f) Persistence diagram of $f$. 
(g) Morse--Smale complex of $f$. (h) A vector field with its critical points: source spirals in red and saddles in yellow. (i) A tensor field visualized by its eigenvector field with degenerate points: trisectors in pink and wedges in white. }
\label{fig:topo-descriptors}
\vspace{-4mm}
\end{figure}

\para{Topological descriptors.}
{\tool} supports a wide range of topological descriptors across multiple data types; see the supplement for their definitions and \cref{fig:topo-descriptors} for representative examples.

For scalar fields, {\tool} computes critical points, persistence diagrams, merge trees, contour trees, and Morse--Smale segmentations.
Critical points are locations where the gradient vanishes; a contour tree (or merge tree) captures how connected components of level sets (or sublevel sets) appear, split, merge, and disappear as the scalar value increases; and a (sublevel-set) persistence diagram records the birth and death of topological features in the sublevel sets and quantifies their significance. A Morse--Smale segmentation partitions the domain into regions whose gradient flows connect pairs of critical points. 

For vector fields, {\tool} identifies critical points. For tensor fields, it computes degenerate points as well as eigenvalue and eigenvector partitions: degenerate points occur where eigenvalues coincide, while eigenvalue/eigenvector partitions segment the domain according to qualitative changes in how tensors behave as a linear operator; see \cref{fig:tensors}(b)-(c). 

{\tool} also supports tracking critical points (scalar and vector fields) and degenerate points (tensor fields) over time via several user-selectable optimal-transport-based algorithms.
All descriptors can be visualized independently or overlaid on the original data.

\para{Derived attributes.}
{\tool} includes several computations on derived fields, including persistence simplification for scalar fields, gradient computation for scalar and vector fields, and magnitude computation for vector and tensor fields. 

%% file: sec-method.tex
\section{Implementation with Failure-Mode Mitigation}
\label{sec:method}

\subsection{Architecture and Implementation}
\label{sec:method-implementation}

\new{We now discuss selected implementation details. Additional details, including those concerning the front end and code generation, are provided in the supplement.}

\subsubsection{Computation Engine}

\para{Input files.} {\tool} currently supports structured data stored in the VTK image format (.vti). The user must specify which arrays in the dataset correspond to each field. Time-varying datasets can be loaded by providing {\tool} with the name of a directory containing sequentially numbered files, where array names must remain consistent across all time steps.

\para{Data types.} Internally, {\tool} operates on a variety of data types, such as scalar fields, vector fields, tensor fields, and various topological descriptors. To ensure consistency across operations, we implement an internal type system. Specifically, we define a custom abstract class called \texttt{NodeData} that is extended to implement each data type. When applicable, data types include logic describing how they should be exported. Some data types are \textit{embedding} types (e.g., contour tree embeddings), which store the information required to generate visualizations, such as the data to render and associated color mappings. These type definitions also include routines specifying how the data should be visualized.

\begin{figure}[!ht]
\vspace{-2mm}
\begin{center}
\includegraphics[width=0.8\linewidth]{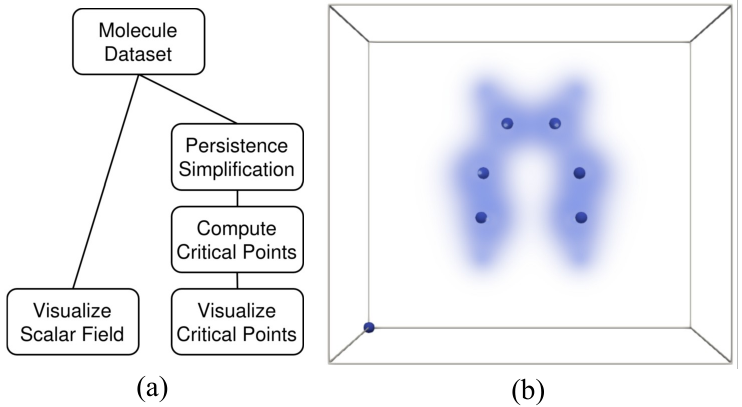}
\vspace{-5mm}
\caption{(a) Node tree diagram for showing the molecule dataset with local minima after persistence simplification. (b) Final visualization from (a) with the minima in blue.}
\label{fig:molecule}
\vspace{-6mm}
\end{center}
\end{figure}

\para{Node tree.} At the core of each visualization is a custom data structure that we call the \textit{node tree}. The node tree is a rooted tree that represents an abstraction of a workflow. Each node corresponds to a single operation that transforms input data into output data. The root node is the only exception, as it performs data loading rather than a transformation. During execution, the root node loads data from disk and passes it to its children. Each subsequent node receives data from its parent, applies its transformation, and then forwards the resulting data to its children. A simple node tree example is shown in \cref{fig:molecule}(a) with its resulting visualization in (b). We further describe how nodes are implemented in the supplement.

\para{LLM Tools.} LLMs interact with the {\tool} backend through five primary tools, which can be provided in the OpenAI or Anthropic API format or through MCP. The \texttt{create\_node} tool adds a node to the node tree and accepts the node's input parameters as a dictionary. To obtain information about a specific node type, including the parameters required to instantiate it, the LLM can call \texttt{node\_info}. When \texttt{create\_node} is called, it validates the supplied parameters and verifies that the node's input data type matches its parent node's output type. If validation succeeds, the node is created and its ID is returned to the caller. Otherwise, the node is not created, and the caller receives actionable feedback. Two analogous tools, \texttt{export\_node} and \texttt{export\_info}, support saving data to disk.

Finally, \texttt{visualize\_embeddings} accepts a list of nodes whose outputs are embedding types. It triggers computations in the node tree and renders the resulting embeddings in the visualization window. The visualization is also returned to the orchestrator, allowing it to inspect the results and make adjustments when necessary.

\subsubsection{Reliability-Centered Two-Agent Architecture}

\para{Orchestrator agent.} Our system implements two agents. The primary agent, called the \textit{orchestrator}, constructs the node tree using the available tools and invokes \texttt{visualize\_embeddings} to produce a visualization. All interactions between the user and {\tool} occur through the orchestrator.

\para{Verifier agent.} The second agent, called the \textit{verifier}, is responsible for validating proposed visualizations. Whenever the orchestrator invokes \texttt{visualize\_embeddings}, the verifier checks that the visualization aligns with the user’s intent and that no errors are present. To do so, we provide the verifier with the full chat history along with Python code that reconstructs the node tree. For each node, we include descriptions of its parameters and other relevant information (e.g., whether persistence simplification is required).

The verifier answers a series of yes/no questions that assess different aspects of the visualization’s correctness, and returns answers through a single tool. Using a tool ensures that responses are constrained to boolean values, enabling deterministic downstream logic. Each question also includes a string parameter that allows the verifier to explain its reasoning.

When the orchestrator calls \texttt{visualize\_embeddings}, the visualization is generated only if the verifier reports no issues. Otherwise, visualization is blocked and the verifier’s feedback is returned to the orchestrator. The specific validation questions are described in \cref{sec:method-failure-modes}, with the exact prompts provided in the supplement.

\subsection{Mitigation of Failure Modes}
\label{sec:method-failure-modes}

We first describe a general guardrail strategy for mitigating multiple types of failure modes. 
Guardrails constrain agent behavior to ensure reliability. 
We then enumerate failure modes and explain how each is addressed using these guardrails, along with additional mitigation strategies.

\subsubsection{Guardrail}

A common strategy we use to prevent failure modes is what we call a \emph{guardrail}. In our context, a guardrail constrains the orchestrator to a feasible action space and requires the verifier to approve each workflow prior to execution. For example, a guardrail can require the orchestrator to ask the user whether persistence simplification should be applied before computing critical points.

Guardrails influence behavior in three ways. First, a description of the required behavior is appended to the tool description associated with the node type. Second, the generated tool includes a boolean parameter that the orchestrator must set to \texttt{true} only if it has performed the required behavior. If this parameter is set to \texttt{false}, the node may be created differently than requested or its creation may be canceled, and the orchestrator receives feedback. Finally, the verifier is informed of all guardrails and checks whether they have been satisfied. If the verifier detects a violation, its explanation is returned to the orchestrator.

Common guardrails (e.g., enforcing default parameter values) are generated automatically. We also implement a \texttt{Guardrail} class that enables custom guardrails to be instantiated  for individual node types as \emph{guardrail instances}. These include guardrails that constrain scalar field embedding, eigenvector partition embedding, isocontour embedding, and persistence simplification. A complete list of node-specific guardrails is provided in the supplement.

\subsubsection{Mitigation Strategies for Failure Modes}

We enumerate each failure mode and describe how it is mitigated. As illustrated in \cref{fig:teaser}, the failure modes \inlineicon{f1.pdf} through \inlineicon{f6.pdf} may arise at different stages of the system workflow and evaluation loop, where safeguards \inlineicon{guardrail.pdf}---including guardrails and additional mitigation strategies---are applied to address them.

\para{\ClarificationFailureF. Clarification failure \inlineicon{f1.pdf}.} \new{To address underspecified or ambiguous requests,} we implement guardrails that prevent the creation of certain nodes until the user has provided the required information. This information may include parameter values (e.g., the backend used for feature tracking) or details about the intended workflow (e.g., whether persistence simplification should be applied).

\para{\ConfusionOfCapabilitiesF. Confusion of capabilities \inlineicon{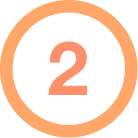}.} When {\tool} starts, we provide the orchestrator with a global prompt that lists its available capabilities. The orchestrator is explicitly instructed not to attempt tasks outside these capabilities. The prompt also encourages the orchestrator to declare a task infeasible if it is not supported. The full global prompt is provided in the supplement.

\para{\BadParameterValuesF. Invalid \new{operation} \inlineicon{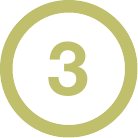}.} \new{Since the orchestrator is restricted to valid operations, this failure mode can arise only from parameter choices, in two ways.} First, the orchestrator may supply an invalid parameter value, such as a negative persistence threshold. Parameter values are validated during node creation. If an invalid value is detected, node creation is blocked, and the orchestrator receives feedback explaining the error.

Second, the orchestrator may select parameter values that lead to a poor-quality visualization. To mitigate this issue, we provide a reasonable default value for each parameter. Guardrails enforce the use of these default values unless the user explicitly overrides them.

\para{\InvalidWorkflowF. Invalid workflow \inlineicon{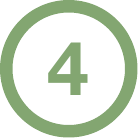}.} This failure mode can occur in two ways. First, the orchestrator may attempt to construct a semantically invalid workflow (e.g., applying persistence simplification to a vector field). We prevent such errors using our type system. Each node declares an input type and an output type, and operations can only be composed when these types are compatible.
Second, invalid workflows may arise during visualization, such as attempting to overlay incompatible visual elements (e.g., overlapping scalar fields). To address this issue, we define rules for each embedding type specifying which other embedding types may be visualized simultaneously \new{and enforce them deterministically}. If a visualization violates these rules, it is rejected and the orchestrator receives feedback.

\para{\IntentionsNotSatisfiedF. Goal misalignment \inlineicon{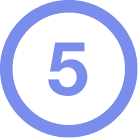}.} We check whether the user’s intentions are satisfied using feedback from the verifier. In addition, some user requests may be technically ambiguous but have a clear practical interpretation. For example, if the user asks the system to ``apply persistence simplification and compute the critical points,'' the likely intent is to compute the critical points of the simplified scalar field, even though the request could technically be interpreted as two independent operations. We resolve specific anticipated ambiguities using targeted guardrails that enforce the intended workflow.

\para{\ErroneousExecutionF. Erroneous execution \inlineicon{f6.pdf}.} Once the node tree is constructed, its execution is fully deterministic and produces consistent results, assuming correct implementation of the evaluation backend. Consequently, the correctness of the final visualization follows directly from the correctness of the constructed node tree.

%% file: sec-case-studies.tex
\section{Case Studies}
\label{sec:case-studies}

We present a series of case studies that demonstrate the capabilities of {\tool} in conversational workflow automation, highlighting its ability to operate on scalar, vector, and tensor field data across a number of topological data analysis and visualization tasks. Descriptions of the datasets, \new{additional case studies}, and full transcripts are given in the supplement.

\subsection{Analysis and Visualization of Scalar Fields}
\label{sec:scalar-fields}

\begin{figure}[!ht]
\centering 
\vspace{-3mm}
\includegraphics[width=0.8\linewidth]{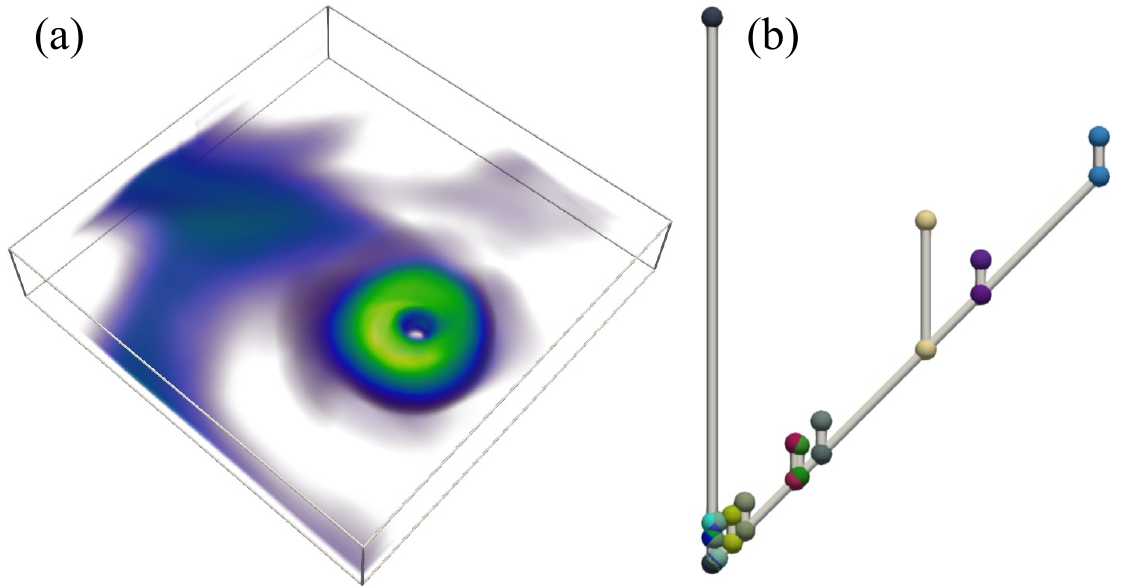}
\vspace{-3mm}
\caption{Computing the simplified persistence diagram with {\tool}. (a) Volume rendering of the Hurricane Isabel dataset with an automatically generated transfer function using the Kindlmann colormap, shown alongside (b) a simplified persistence diagram.}
\label{fig:persistence}
\vspace{-3mm}
\end{figure}

\para{Compute and simplify persistence diagrams.} We demonstrate {\tool}’s ability to compute and visualize the Hurricane Isabel wind speed dataset and its persistence diagram following persistence simplification. Persistence diagrams highlight prominent topological features and are commonly used as a starting point for analysis. However, generating them typically requires a multi-step workflow involving data loading, simplification, diagram computation, and visualization customization. With {\tool}, this entire process can be performed through natural language interaction. We provide an excerpt of the conversation, with the user’s prompt in \usercolor{green} and the agent’s response in \agentcolor{blue}, where the resulting visualizations are shown in \cref{fig:persistence}: 
\noindent\myuser{Please load and visualize Isabel.vti with the Kindlmann colormap.} 
\myagent{Is ... [it] a scalar field?}
\myuser{It is a scalar field.}
\myagent{Loaded Isabel.vti ... and visualized it.}
\myuser{Now please visualize the simplified persistence diagram.}
\myagent{What persistence-simplification threshold should I use?}
\myuser{Please use a relative threshold of 0.04.}
\myagent{Applied ... persistence-simplification ... then computed and visualized the ... persistence diagram.}

\begin{figure}[!ht]
\vspace{-3mm}
\begin{center}
\includegraphics[width=0.8\linewidth]{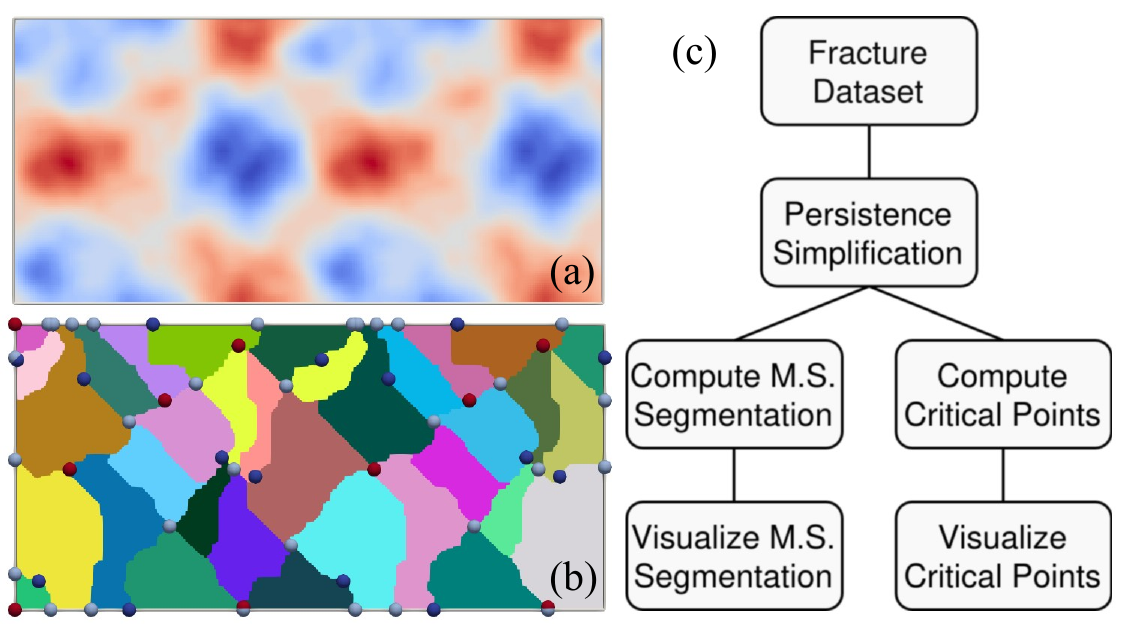}
\vspace{-3mm}
\caption{(a) Fracture dataset. (b) Simplified Morse–-Smale segmentation with its critical points: maxima in red, saddles in light blue, and minima in blue. (c) Node tree diagram.}
\label{fig:case-studies-fracture}
\vspace{-8mm}
\end{center}
\end{figure}

\para{Extract Morse--Smale segmentation.} 
We extract the Morse--Smale (MS) segmentation of the surface of a jagged fractured stone. The dataset is a 2D scalar field in which each value represents the height of the stone at that point. By analyzing the MS complex, we isolate distinct peaks and valleys to study the structure of the fracture. As in previous examples, {\tool} translates a simple natural-language prompt into a multi-stage pipeline that produces several visualizations. The final result is shown in \cref{fig:case-studies-fracture}.

\begin{figure}[!ht]
\centering
\vspace{-2mm}
\includegraphics[width=0.7\linewidth]{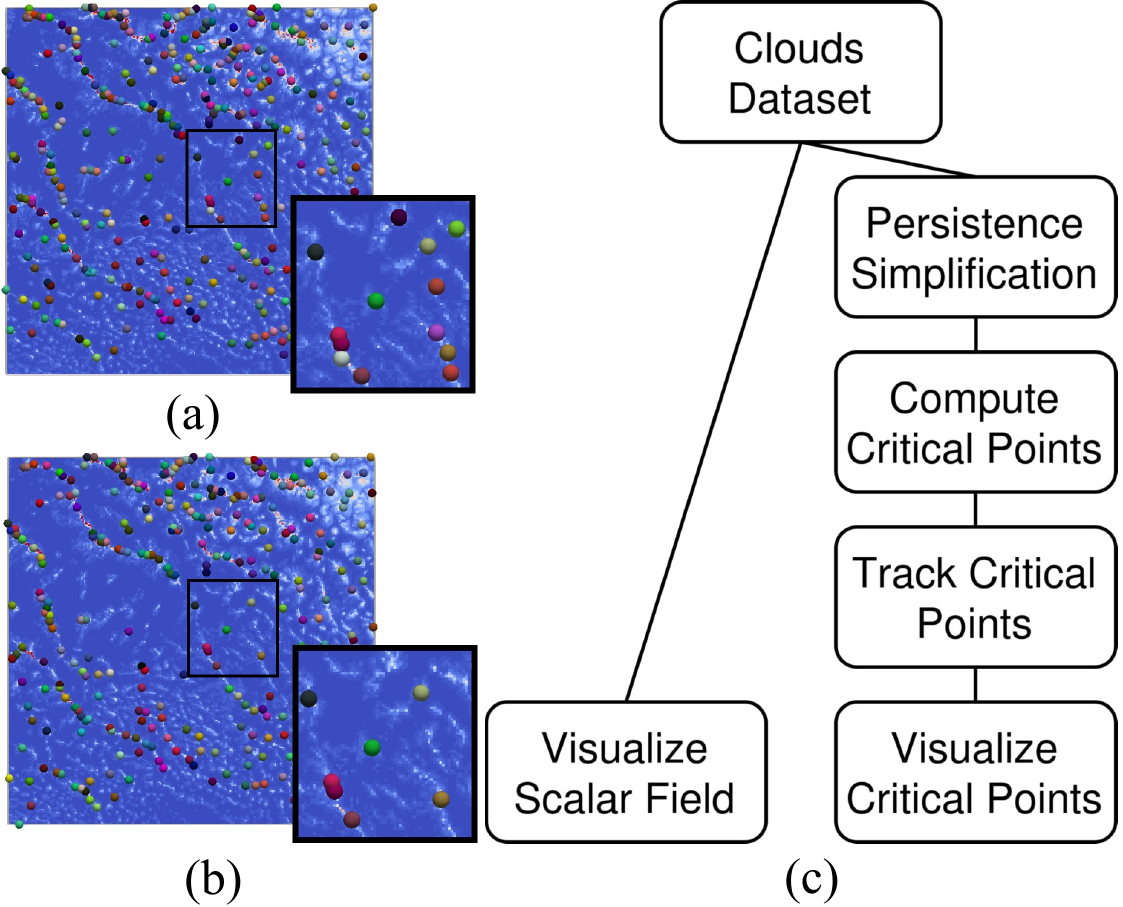}
\vspace{-4mm}
\caption{Critical point tracking of the Cloud dataset. (a)–(b) Visualizations of the first two time steps highlighting the tracked local maxima. (c) Node-tree diagram for the cloud-tracking workflow.}
\vspace{-2mm}
\label{fig:clouds}
\end{figure}

\para{Critical point tracking.} 
We now demonstrate feature tracking with {\tool}, a workflow that typically involves three distinct stages---feature extraction, tracking, and visualization---each of which often requires separate frameworks or libraries that must be manually integrated. For this example, we track the motion of critical points, specifically local maxima associated with cloud optical fields of low-level clouds, using satellite images from Li et al.~\cite{li2025tracking}. Clouds captured in satellite data are complex, time-varying phenomena involving numerous events, including the formation, dissipation, merging, and splitting of cloud systems. {\tool} enables users to perform all three stages of feature tracking seamlessly using only natural-language prompts. The final visualization and node-tree diagram are shown in \cref{fig:clouds}.

\subsection{Analysis and Visualization of Vector Fields}
\label{sec:vector-fields}

\para{Vector field visualization.} We visualize the phase space of the Duffing equation, which is an ODE that describes the motion of a damped oscillator in 1D. The phase space is a vector field that represents the possible states of the equation at a given time step. The topology of such fields is of interest to scientists as the critical points correspond to equilibrium states, while the critical types determine the stability. The final visualization is shown in \cref{fig:topo-descriptors}(h).

\begin{figure}[!ht]
\centering
\vspace{-3mm}
\includegraphics[width=0.8\linewidth]{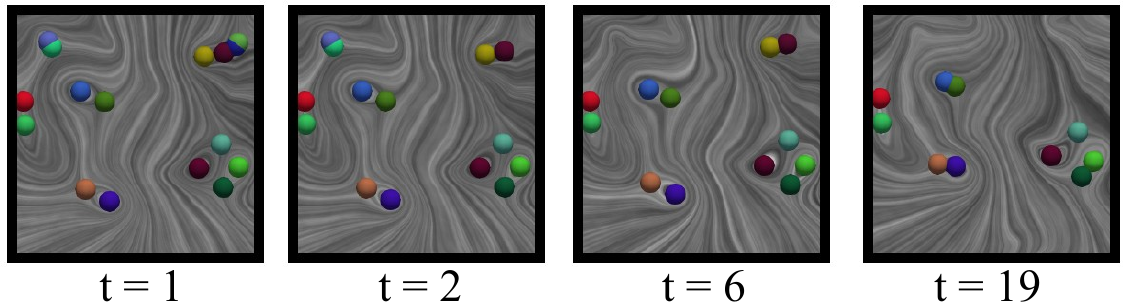}
\vspace{-3mm}
\caption{Zoomed-in views of critical-point tracking using the Cylinder dataset. Four time steps are shown at key moments when critical points appear or disappear, with points rendered in consistent colors.}
\label{fig:cylinder}
\vspace{-2mm}
\end{figure}

\para{Critical point tracking.} We visualize a time-varying heated-cylinder flow and track the locations of its critical points (such as centers and saddles). This dataset describes how a viscous fluid flow travels around a cylindrical object. In this setting, the critical points correspond to real features, such as vortices, and tracking their locations allows scientists to understand how features travel and interact over time. As noted earlier, feature tracking is often complex, and {\tool} assists in constructing multi-stage tracking workflows. The tracking results are shown in \cref{fig:cylinder}. The results demonstrate that the tracking remains consistent across time steps and is robust to the appearance and disappearance of critical points. 

\subsection{Analysis and Visualization of Tensor Fields}
\label{sec:tensor-fields}

\begin{figure}[!ht]
\begin{center}
\includegraphics[width=0.8\linewidth]{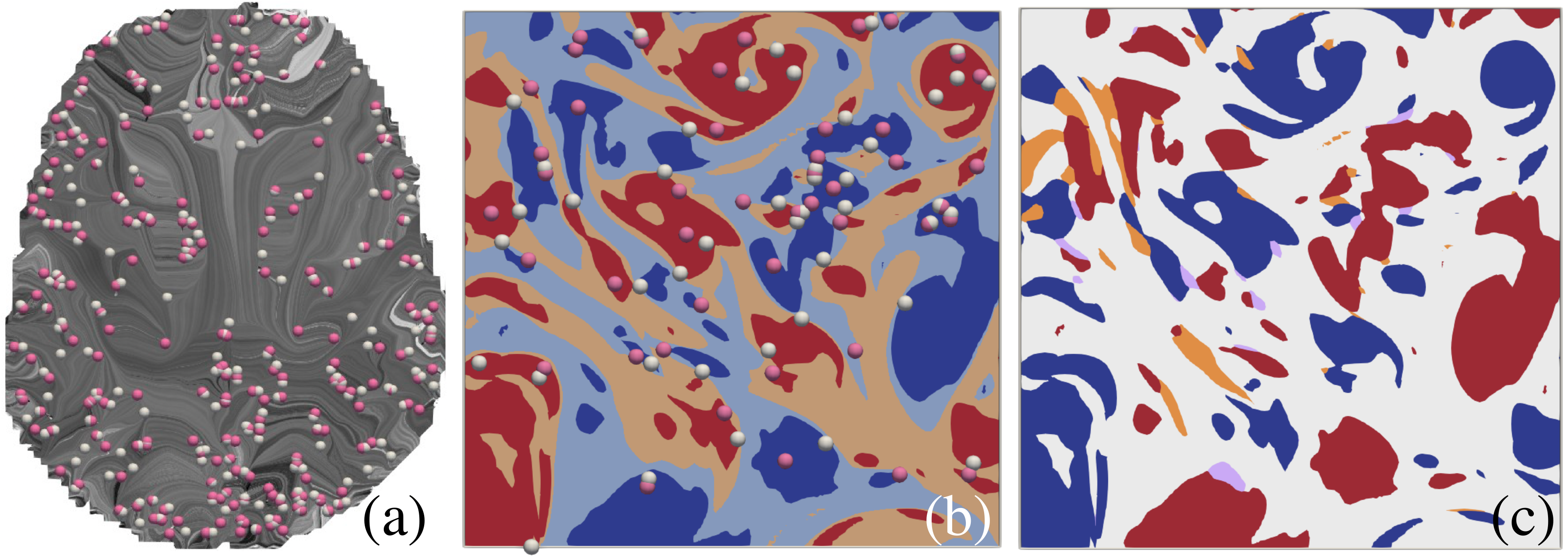}
\vspace{-3mm}
\caption{Tensor field visualizations produced by {\tool}. (a) HyperLIC visualization of the Brain dataset along with its degenerate points: trisectors in white, wedges in pink. (b) Eigenvector and (c) eigenvalue partition of the Ocean dataset.}
\label{fig:tensors}
\end{center}
\vspace{-10mm}
\end{figure}

\para{Symmetric tensor field visualization.} We generate a hyperLIC visualization of a symmetric tensor field derived from a brain MRI scan, together with its degenerate points. In tensor fields, degenerate points mark locations where tensor eigenvalues coincide, indicating transitions in anisotropy and changes in the underlying structural organization. In the context of brain data, these points can highlight boundaries between tissue types, regions of directional uncertainty, or structural features related to neural pathways. Despite their importance, support for symmetric tensor-field visualization is generally limited. {\tool} makes this otherwise specialized visualization accessible through natural-language interaction. The final visualization appears in \cref{fig:tensors}(a).

\para{Asymmetric tensor field visualization.} We visualize the eigenvector and eigenvalue partitions of an asymmetric tensor field from an Ocean dataset. The tensor field represents the gradient of the ocean flow. {\tool} supports multiple visual encodings for tensor fields. The eigenvector partition of the tensor field in \cref{fig:tensors}(b) highlights flow rotation direction, and the strength of fluid flow relative to anisotropic flow behavior. The eigenvalue partition in (c) demonstrates how each tensor behaves as a linear operator. 

%% file: sec-results.tex
\section{Experimental Evaluation}
\label{sec:results}

We evaluate the reliability of {\tool} across a diverse range of tasks with and without its safeguards enabled. For comparison, we assess the ability of Claude Code and OpenAI Codex to perform the same tasks, both with and without access to the {\tool} source code (allowing it to utilize the computation engine).

\subsection{Experiment Overview}

\para{Agent setup.} We evaluate {\tool} using a custom harness that supplies prompts and captures the outputs. To support {\tool}'s ability to ask follow-up questions, we implement a separate \textit{tester} agent that simulates a user in multi-turn conversations. The tester can provide information stored as key--value pairs. It has access to the keys but not the values and therefore cannot determine whether {\tool} has made an error (see the supplement for details). We evaluate {\tool} both with and without its safeguards enabled. Specifically, we enable or disable selected instructions in the global prompt, all guardrails (e.g., parameter-validity, node-tree-compatibility, and visualization checks), and the verifier agent.

We run Claude Code and Codex in headless mode within a Docker environment containing all libraries required by {\tool}, \new{so neither agent needs to implement standard topological computations from scratch}. We evaluate both agents with and without access to {\tool}. When access is provided, we supply the complete source code, install the {\tool} library in the local Python environment, and \new{provide detailed documentation in a text file}.

We run {\tool} with GPT-5.6 Terra (reasoning effort: \texttt{low}) and Claude Sonnet 5 (effort: \texttt{low}). We selected these lower-cost base models and reasoning-effort settings to limit inference costs, as preliminary experiments indicated that {\tool} performs well under these settings. To distinguish the effects of agentic harnesses from base models, we evaluate Codex and Claude Code using the same base models and reasoning effort settings.

\para{Tasks and prompts.} We evaluate 10 tasks: eight supported by {\tool} (i.e., \emph{feasible}) and two intentionally \emph{infeasible} tasks that test its handling of unsupported requests. One infeasible task is well-defined but unsupported (visualizing a persistence barcode), while the other is semantically invalid (simplifying a scalar field according to its gradient). All supported tasks require multistep workflows, and seven involve visualizing multiple elements.

We generate 10 prompts per task. For feasible tasks, three prompts give complete information, four give partial information, and three give minimal. For Claude Code and Codex, we evaluate the three complete prompts for each supported task (24 total), augmented to include information about its environment (e.g., available libraries). The supplement provides all tasks and prompts.

We define a single evaluation of a prompt as a \emph{trial}. For each {\tool} setting, we run each of the 100 prompts 10 times, yielding 1{,}000 trials. Due to time and cost constraints, we run each of the 24 prompts four times for each Claude Code and Codex setting, yielding 96 trials per setting.

\para{Evaluation.} For each trial, we record input and output token counts and wall-clock time, including both model response and computation time. For {\tool}, we report these metrics only for feasible tasks, as infeasible tasks require fewer resources. \new{We provide the metrics for infeasible tasks in the supplement.}

When applicable, we use the automated scoring system from SciVisAgentBench. It assesses a subset of the computed data rather than the visualizations, and assigns scores from 0 to 10, with 10 indicating perfection. We list specifics in the supplement.

For feasible {\tool} trials that produce a visualization, we compare the resulting node tree with a ground-truth tree. A trial is correct if the two trees yield identical visualizations. For infeasible tasks, we manually score a trial as correct if {\tool} determines that the request cannot be completed, including cases in which it initially proceeds but self-corrects. For Codex and Claude Code, we manually assess the generated code and visualizations against each prompt's computational and visualization requirements. We permit many potential implementations or visualizations that broadly satisfy the requirements. Specifically, we develop rubrics based on \new{the individual items requested in each prompt} and score the results using a custom interface. We also record minor errors, such as mishandled edge cases or ineffective visual encodings. The supplement provides a link to the scoring interface, complete scored results, and additional evaluation details. Although the evaluation criteria are not identical across Claude Code, Codex, and {\tool}, we believe the accuracy comparison remains meaningful. \new{The criteria for {\tool} are stricter: only one code output is accepted as correct, whereas outputs from Codex and Claude Code are considered correct if they broadly fulfill the prompt}.

\begin{table}
\centering
\setlength{\tabcolsep}{3pt}
\resizebox{\columnwidth}{!}{%
\begin{tabular}{l|cccccc} \hline
{\tool} model & time$\downarrow$ & I.T.$\downarrow$ & O.T.$\downarrow$ & B.S.$\uparrow$ & I.F. Acc.$\uparrow$ & F. Acc.$\uparrow$ \\ \hline
GPT & \textbf{23.0} & 32.8K & 744 & \textbf{10.0} & 90.5\% & \textbf{99.0\%}\\
GPT (control) & 23.2 & \textbf{16.0K} & \textbf{533} & 7.9 & 0.0\% & 38.2\%\\
Claude & 26.3 & 59.3K & 1.4K & 9.9 & \textbf{100.0\%} & 98.6\%\\
Claude (control) & 28.3 & 35.6K & 1.4K & 8.5 & 0.5\% & 40.6\%\\
\hline
Coding Agent & time$\downarrow$ & I.T.$\downarrow$ & O.T.$\downarrow$ & B.S.$\uparrow$ & N.M.E.$\uparrow$ & Acc.$\uparrow$ \\ \hline
Codex & 81.1 & 592.4K & 5.5K & 6.6 & 31.2\% & 8.3\%\\
Codex (T.P.) & 69.0 & 533.7K & 3.3K & \textbf{10.0} & \textbf{92.7\%} & 64.6\%\\
Claude Code & 265.7 & 1.9M & 14.5K & 5.9 & 21.9\% & 2.1\%\\
Claude Code (T.P.) & 93.3 & 567.6K & 4.4K & 9.9 & 90.6\% & \textbf{77.1\%}\\
\hline
\end{tabular}
}
\caption{Average runtime (including rendering), token usage, and accuracy for {\tool},
Claude Code, and Codex. GPT (control) and Claude (control) denote
{\tool} using GPT and Claude, respectively, with safeguards disabled.
Codex (T.P.) and Claude Code (T.P.) denote coding agents with access
to {\tool}. Time: average total runtime, including computation, in
seconds; I.T./O.T.: average input/output tokens; B.S.: average
benchmark score; I.F.Acc./F.Acc.: {\tool} accuracy on infeasible/feasible
tasks; N.M.E.: percentage of coding-agent trials with no major errors;
Acc.: percentage of coding-agent trials with no errors. For {\tool},
runtime and token counts are averaged only over feasible tasks, as
infeasible tasks require fewer resources.}
\label{tab:agent-results}
\vspace{-8mm}
\end{table}

\subsection{Analysis of Evaluation Results}


\subsubsection{{\tool} Performance} 

\begin{highlight}
\para{Key takeaway:} {\tool} has a high accuracy on feasible and infeasible tasks, and achieves near-perfect benchmark scores.
\end{highlight}

In \cref{tab:agent-results}, we can see that {\tool} achieves high accuracy while using a modest number of tokens. On infeasible tasks, {\tool} using GPT succeeds in 90.5\% of trials. In the unsuccessful simplification trials, it applies standard persistence simplification to the domain; in the unsuccessful persistence-barcode trials, it displays a persistence diagram. The Claude backend achieves perfect accuracy on infeasible tasks.

On feasible tasks, {\tool} (GPT) succeeds in 792 of 800 trials (99.0\%). Of the eight failures, six involve assuming a parameter value rather than asking for it. In one remaining failure, it visualizes a scalar field multiple times in one visualization. In the other, it asks for information clearly provided in the documentation. The Claude backend achieves comparable performance, succeeding in 98.6\% of feasible-task trials. Most failures arise from prompts asking the system to ``visualize X and then visualize Y.'' Claude sometimes creates separate visualizations despite a guardrail requiring a single visualization. The remaining failures are similar to those of GPT. \new{Accuracy is comparable across levels of prompt specificity, while more specific prompts require fewer tokens. Details are provided in the supplement.}

\subsubsection{Control Trials}

\begin{highlight}
\para{Key takeaway:} With its safeguards disabled, {\tool} exhibits very poor accuracy overall.
\end{highlight}

\cref{tab:agent-results} reports the performance of {\tool} with safeguards disabled. Accuracy decreases substantially, showing the effectiveness of the safeguards. On infeasible tasks, {\tool} generally produces an alternative output rather than recognizing that the task is infeasible. On feasible tasks, the most common failure is not requesting information. Accordingly, in the control trials, {\tool} performs best on fully specified prompts, achieving success rates on feasible tasks of 50.8\% with GPT and 61.3\% with Claude. These results show the importance of including vague and underspecified prompts in visualization benchmarks. \new{We also see relatively high benchmark scores, but low accuracy overall. This result implies that, often, data is correctly computed yet incorrectly visualized.}

\subsubsection{Comparison with Coding Agents}

\begin{highlight}
\para{Key takeaway:} Compared with Codex and Claude Code, {\tool} achieves a higher accuracy in less time on supported tasks while using substantially fewer tokens.
\end{highlight}

On tasks supported by {\tool}, Claude Code and Codex both make substantial errors while requiring considerably more time and tokens. \new{Admittedly, this comparison is limited to supported tasks; general coding agents can also perform tasks beyond {\tool}'s capabilities. Providing these agents with the {\tool} source code as a library} substantially improves their performance in our tests. However, when accounting for visual design decisions such as camera placement, their accuracy remains lower than that of the standard {\tool} configuration \new{on supported tasks}, and they still require more time and tokens. Thus, {\tool} and general coding agents offer complementary strengths: {\tool} delivers superior performance \new{on supported tasks}, while general coding agents \new{could potentially extend its capabilities. Similar to the control trials, we see high benchmark scores yet low accuracy. This trend is due to data that is correctly computed yet incorrectly visualized, or data that is incorrectly computed, yet is similar to the ground truth.}

%% file: sec-expert.tex
\section{Expert Feedback}
\label{sec:expert}

We conducted two informal 90-minute expert feedback sessions with four domain experts (E1–E4) who varied in their familiarity with topological data analysis.
E1–E3 participated in the first session, and E4 in the second. E1 and E2 are third- and fourth-year doctoral students in chemistry, while E3 is a tenured chemistry professor with multiple publications applying topological analysis to chemical data. E4 is a tenured professor in mechanical engineering. All participants regularly (E1, E3) or occasionally (E2, E4) engage with topological features in their work: the chemists (E1-E3) examine topological patterns in chemical data to evaluate their physical significance, and the mechanical engineer (E4) uses topology to characterize porous material structures.

Each participant received a 20-minute demonstration of {\tool}, illustrating its utility across several scientific domains, followed by a semi-structured interview. The semi-structured format kept the discussion informal and conversational, allowing us to probe participants’ workflows as well as the needs and pain points they encounter when applying topology. During the interviews, we generated additional visualizations in response to participant questions. 

\para{Rapid prototyping and exploratory analysis.} 
Overall, the experts thought that {\tool} could be very useful as an exploratory tool. The general consensus was that {\tool} would be useful for domain scientists with some topological knowledge who wanted to rapidly prototype an analysis and visualization pipeline without having to learn and connect new tools. The experts expressed that in exploratory research, one does not always know what they are looking for, and the rapid prototyping of {\tool} is helpful. E3 remarked: ``One of the benefits of this is a lot of times you're not necessarily sure what you want to visualize until you've explored the data more.''  Yet, writing code and learning new tools often presents a roadblock to rapid exploratory analysis. E1 noted that, if they were asked to compute a new topological descriptor: ``I [wouldn't] even know what [that] descriptor is...let alone the Python library that actually accepts my form of data and will calculate some quantity that's formatted in the correct way.'' Expanding on this claim, E2 noted topological tools are often poorly documented and difficult to integrate into existing workflows. The experts thus appreciated the ability to rapidly produce different kinds of novel topological visualizations from natural language.

\para{Workflow automation and extensibility.}
The experts also valued {\tool}’s workflow automation and extensibility. They emphasized the importance of exporting scripts for execution across data batches; E1 suggested that without this capability, {\tool} would have reduced usefulness for their work.

They were also positive about the conversational nature of {\tool}. E3 and E4 remarked that the questions posed by {\tool} can prompt users to be intentional about parameter choices. E4 noted: ``It makes you think: are you sure you want to do this?'' E3 suggested that, by requiring explicit parameter choices, the system may help users avoid incorrectly assuming that patterns in the data are physically meaningful.

\para{Educational potential.}~E4 also identified educational potential, suggesting that {\tool} could demonstrate LLM-assisted scientific analysis while providing an introduction to interdisciplinary research. E4 emphasized that ``interdisciplinary research is what needs to happen these days,'' noting that, ``a mechanical engineer might say, `I love what [they] are doing [with topology].''' 

The feedback also revealed several areas for improvement, most of which could be addressed  by adding new features to {\tool}. 

\para{Tool delivery mechanism.}
All four experts asked about the ease of accessing {\tool}, whether through local installation or by remotely connecting to a server hosting the system. They noted that if {\tool} were difficult to install, it could introduce challenges similar to those associated with learning new frameworks, thereby diminishing its usability benefits. 
In response, we repackaged {\tool} as a Python library that can be installed using \texttt{pip} in any environment with ParaView and the Topology Toolkit. If such an environment is active, {\tool} can be run by simply entering \texttt{TopoPilot} in a command line.


\para{Data conversion capabilities.}~E1–E3 raised questions about preprocessing data to make it compatible with both {\tool} and their in-house analysis tools, noting that it would be valuable to support data conversion utilities in the conversational interface. E3 noted that: ``the preprocessing piece is one of the bigger hurdles for us.'' Such capabilities could be easily incorporated into {\tool} by connecting the backend to a collection of data conversion scripts. Data preprocessing and file format conversions could be handled seamlessly by introducing new node types into the node tree.

\para{Large datasets.}~Both E1 and E4 raised questions about {\tool}’s ability to load and process very large datasets that do not fit in memory, which is a current limitation of the system. While handling large datasets is inherently challenging, we believe it is feasible to develop a cluster-compatible version of {\tool} that would allow users to prototype visualizations and analyses on smaller data subsets before applying the full pipeline to full datasets.

\para{Trustworthiness of output.}~E3 and E4 asked about the trustworthiness of {\tool}’s outputs. Because {\tool} operates as a black box, users may find it difficult to verify what was produced, which could produce faulty interpretations. E3 noted: ``whenever you make something really black box, people do a bunch of calculations that they don't understand. They over-interpret the data because they over-interpret what they got out of the black box.'' E3 clarified that this limitation is inherent to any black-box tool. They also appreciated that {\tool} can export its workflow as a script, which helps individuals to understand the underlying process. 

We noted that our node-tree architecture helps curate workflows in a controlled manner with known tools and libraries, thereby minimizing hallucinations. 
E1 suggested that exporting a workflow diagram would further clarify the system’s operations (see \cref{fig:case-studies-fracture} as an example), while E3 and E4 recommended developing standard benchmarks to help establish confidence in {\tool}’s trustworthiness. E4 remarked: ``[if there was] a specific small benchmark for different categories, I would trust it.'' 

In response to this feedback, we redesigned our code export tool to generate a clear sequence of function calls representing the workflow's atomic operations, making workflow validation simple.

\para{Minor additions.} The experts also proposed smaller features to enhance {\tool}’s utility. E2 and E4 noted that their work extends beyond visualization and expressed interest in using {\tool} to perform computations and downstream analyses. They appreciated the feature-tracking workflows and suggested enabling direct export of these results for further analysis. In response, we gave {\tool} the ability to export the results of select computations. 

E4 further emphasized that expanding the feature-tracking capabilities would be  beneficial and recommended adding more support for flow analysis, such as tools for tracing particle trajectories. E1 suggested implementing more intuitive mechanisms for adjusting parameters, such as slider widgets for variables like persistence simplification thresholds. Finally, E4 suggested having more labels so that diagrams could be interpreted by other users, stating that ``this requires a lot of expertise to understand what [you] are looking at ... Maybe a little bit more labeling ... would be helpful.''

%% file: sec-limitations.tex
\section{Limitations and Future Work}
\label{sec:limitations}

\para{Limitations.}
While {\tool} provides a flexible foundation for extension, it has several limitations. The node-tree architecture simplifies extensibility but, in the absence of code generation, restricts expressiveness. For example, if the user requests visualizing the cosine of a scalar field, this could not be fulfilled. This limitation suggests a clear path forward: expanding the node library and introducing controlled forms of code generation (e.g., nodes that apply LLM-generated mathematical expressions, similar to ParaView’s calculator filter), thereby increasing expressiveness while preserving safety and structure. {\tool} may also suggest hallucinated tasks that it cannot execute. Although it typically recognizes such infeasible requests, stronger safeguards are needed to prevent the inference of nonexistent capabilities. \new{Finally, we improve the accuracy of {\tool} by anticipating linguistic ambiguities and potential failures. However, this approach cannot eliminate issues we have not foreseen.}

\para{From assistant to tutor.}
Early expert feedback indicates that {\tool} is well suited for domain scientists familiar with topological analysis but lacking the time or expertise to construct complex workflows. Likewise, its conversational interface makes it accessible to novices, such as students new to topological methods by supporting explanation, method recommendation, and guided workflow construction. Thus, {\tool} functions both as an expert assistant for rapid prototyping and as a tutor that builds intuition through interaction. This dual role has the potential to broaden access to topology-based visualization. A systematic study of {\tool}'s effectiveness as a pedagogical tool is left to future work. More broadly, its architecture offers a foundation for automating a wide range of scientific analysis and visualization pipelines.

%% file: sec-ack.tex
\section*{Acknowledgment}
This work was performed under the auspices of the U.S. Department of Energy (DOE) by Lawrence Livermore National Laboratory under Contract DE-AC52-07NA27344. The work was partially funded by DOE ECRP 51917/SCW1885 and DOE DE-SC0023157. This work is reviewed and released under LLNL-CONF-2017332. We thank Aurora Clark, Joshua Bilsky, Jackson Elowitt, and Pania Newell for feedback. We used OpenAI Codex for refactoring and bug fixing in {\tool}, as well as for most of the development of the grading application and some data processing code. ChatGPT was also used for grammar correction and language polishing.

%% file: appendix-dataset-descriptions.tex
\section{Descriptions of Datasets}
\label{appendix:dataset-descriptions}

We describe the datasets used in the case studies and experiments. The \textbf{Isabel} dataset is derived from a simulation conducted by the National Center for Atmospheric Research and is included in the 2004 IEEE VIS SciVis contest~\cite{scivis2004}. Its original resolution is $500 \times 500 \times 100$; we truncate it to $500 \times 500 \times 90$ to remove regions primarily over land where no data is available. We use the wind-speed field, normalized to $[0,1]$.

The \textbf{Ionization} dataset is derived from the ionization-front simulation of Whalen and Norman~\cite{whalen2008ionization}, featured in the 2008 IEEE VIS SciVis contest~\cite{scivis2008}. We access the data via the contest webpage and use the time-varying velocity field, specifically time steps 100 through $120$ (inclusive). Each field is downsampled by selecting one out of every four points along each coordinate axis, yielding a $64\times$ reduction. At each sampled point, the scalar value is defined as the magnitude of the velocity vector.

The \textbf{QMCPACK} dataset is taken from the SDR bench~\cite{zhao2020sdrbench} and is based on a QMCPACK continuum quantum Monte Carlo performance test \cite{kim2018qmcpack, kent2020qmcpack}. We isolate the 145th orbital (out of 288) and normalize its values to $[0,1]$.

The \textbf{Fracture} dataset is drawn from the Fractures With Variable Roughness and Wettability  collection~\cite{guiltinan2020fractures}, accessed via the Digital Porous Media Portal~\cite{prodanovic2025digital, turhan2024digital}. We use the “top” field from fracture 15 with a fractal dimension of 1.5. To smooth the data, we apply a $7 \times 7$ convolutional filter (without padding), replacing each center pixel with the mean of the values within the filter window.

From our collaborators, the \textbf{Molecule} dataset represents an electron density field, whereas the \textbf{Cloud} dataset represents a cloud optical depth (COD) field derived from satellite imagery of low-level clouds~\cite{li2025tracking}, where higher values typically indicate thicker clouds.

The \textbf{Damped Oscillator}~\cite{Haller11} and \textbf{Heated Cylinder}~\cite{Guenther17} datasets are derived from simulations made available through the Computer Graphics Laboratory at ETH Zurich. The heated-cylinder simulation is performed using the Gerris Flow Solver~\cite{gerrisflowsolver}.

The \textbf{Brain} dataset is obtained from a public MRI dataset by Tian et al.~\cite{tian2022comprehensive}. We use data from patient 23 and extract the diffusion tensor field using the Diffusion Imaging in Python (DIPY) library~\cite{garyfallidis2014dipy}.

The \textbf{Ocean} dataset is obtained from the Global Ocean Physics Reanalysis provided by the EU Copernicus Marine Service~\cite{oceanData}. We compute the tensor field by taking numerical gradients of a flow field constructed from the ``uo'' and ``vo'' components of the daily data (file name: \textsf{cmems\_mod\_glo\_phy\_my\_0.083deg\_P1D-m}). The data is sliced over the ranges $x: 100\text{--}200$, $y: 10\text{--}110$, and $z = 0$ (all inclusive), where $z$ denotes depth. We use data from June 2, 2019.

Because the Ocean dataset is a vector field, we derive an asymmetric tensor field by approximating its gradient using finite differences. For completeness, we briefly summarize the computation of $\partial v_x / \partial x$, with analogous expressions applied to the remaining partial derivatives. Assuming the flow field is defined on an $n \times m$ grid, we evaluate the derivative at a point $p = (p_x, p_y)$ using separate formulas for interior and boundary points. Specifically:

\begin{itemize}[noitemsep]
\item {$1 < p_x < n$:} $\left.\frac{\partial v_x}{\partial x}\right|_p = v_x(p_x+1,p_y) - v_x(p_x-1,p_y)$;
\item {$p_x = 1$:} $\left. \frac{\partial v_x}{\partial x}\right|_p = v_x(2,p_y) - v_x(1,p_y)$;
\item {$p_x = n$:} $\left. \frac{\partial v_x}{\partial x} \right|_p = v_x(n,p_y) - v_x(n-1,p_y)$.
\end{itemize}

%% file: appendix-descriptors.tex
\section{Topological Analysis Supported by {\tool}}
\label{appendix:descriptors}

We briefly describe the topological descriptors for scalar, vector, and tensor fields currently supported by {\tool}. These constitute an initial set of capabilities; because {\tool} is designed with modular, descriptor-agnostic interfaces, it can be readily extended to incorporate additional topological descriptors and analyses.

\subsection{Topological Analysis of Scalar Fields}
A scalar field is a function $f:\X \rightarrow \R$. Assume $\X$ is a compact domain.

\para{Critical points.} The \emph{critical points} of $f$ are points $x \in \X$ where $\nabla f(x) = 0$. Critical points can be either maxima, minima, or saddle points. The critical points provide a high level summary of the topology of the domain and serve as the basis for more complex topological descriptors. We visualize critical points in \cref{fig:topo-descriptors}(a). 

\para{Morse--Smale segmentation.} The \emph{Morse--Smale segmentation} of $f$ partitions the domain according to gradient flow behavior. Every regular (non-critical) point $x$ lies on a unique integral line of the gradient field, which originates at a local minimum $p$ and terminates at a local maximum $q$, following the direction of increasing function value. Let $\alpha$ denote the mapping that assigns each point $x$ to the pair $(p, q)$ corresponding to the origin and destination of its integral line. The Morse--Smale segmentation partitions the domain $\X$ into regions such that two points $x$ and $y$ belong to the same region if and only if $\alpha(x) = \alpha(y)$. See~\cite{maack2023parallel} for details on the definition and computation. We visualize the Morse--Smale segmentation in \cref{fig:topo-descriptors}(g).

\para{Merge tree.} The \emph{merge tree} is defined in terms of the sublevel sets of $f$, denoted $\X_t = \{x \mid f(x) \leq t\}$. It is a rooted tree that encodes how the connected components of $\X_t$ appear and merge as $t$ increases. Each local minimum initiates a new connected component and corresponds to a leaf of the tree. As $t$ increases, components merge at critical points; these merge events occur at saddle points, which form the internal nodes of the tree. The root corresponds to the global maximum, at which all components have merged into a single connected component. We visualize the merge tree in \cref{fig:topo-descriptors}(c).

\para{Contour tree.} The \emph{contour tree} is defined in terms of the level sets (contours) $f^{-1}(t)$. It captures how the connected components of these level sets appear, merge, and split as $t$ varies. Each local minimum and maximum corresponds to a leaf node of the contour tree, while the internal nodes correspond to saddle points. We visualize the contour tree in \cref{fig:topo-descriptors}(d); see~\cite{carr2003computing} for a detailed description of merge and contour trees.

\para{Persistence diagram and simplification.} In our context, we consider a sublevel-set \emph{persistence diagram}~\cite{edelsbrunner2008persistent}, which provides a summary of how the topology of $\X_t$ evolves as $t$ increases by recording the birth and death of topological features. As $t$ grows, new features appear and existing ones disappear. For example, a connected component is \emph{born} at a local minimum and \emph{dies} at a saddle point when it merges with an older component. Similarly, higher-dimensional features (e.g., loops or voids) arise and vanish through interactions of critical points, as described by Morse theory~\cite{milnor1963morse}. Each feature is assigned a birth value $b$ and a death value $d$, and its \emph{persistence} $d - b$ reflects its significance.

The persistence diagram represents these features as a scatter plot with coordinates $(b,d)$. Points farther from the diagonal $b = d$ correspond to persistent (and typically more significant) features, while points near the diagonal often indicate noise or less prominent structures. Building on this, \emph{persistence simplification} removes features whose persistence falls below a user-specified threshold $\varepsilon$, effectively smoothing out topological noise. In practice, descriptors such as the Morse--Smale segmentation, merge tree, contour tree, and persistence diagram are typically analyzed after such simplification. Our current persistence diagram backend only supports orders $0$ and $D-1$, where $D$ is the dimension of the dataset. We visualize persistence simplification applied to a scalar field in \cref{fig:topo-descriptors}(b), along with the corresponding simplified contour tree in \cref{fig:topo-descriptors}(e), and the persistence diagram in \cref{fig:topo-descriptors}(f). See~\cite{zomorodian2004computing, vidal2021progressive} for definitions and computation, and~\cite{tierny2012generalized} for simplification details.

\subsection{Topological Analysis of Vector Fields}

A vector field is a function $v : \X \rightarrow \R^n$ that assigns a vector to every point in the domain.

\para{Critical points.} A \emph{critical point} of $v$ is a point $x \in \X$ where $v(x) = 0$, in contrast to scalar-field critical points defined via vanishing gradients. Critical points characterize the flow behavior of the vector field. Their types are determined by the behavior of the field in a neighborhood of the point (e.g., sources, sinks, saddles). We visualize a vector field together with its critical points in \cref{fig:topo-descriptors}(h). See~\cite{gunther2021introduction} for a more detailed description of vector field topology.

\subsection{Topological Analysis of Tensor Fields}

In our setting, we treat a tensor as a multidimensional array of numbers. Let $\T$ denote a space of tensors of fixed order and dimension. A \emph{tensor field} is a function $T : \X \rightarrow \T$ that assigns a tensor to each point in the domain. The field is \emph{symmetric} if $T(x)$ is symmetric for all $x \in \X$, and \emph{asymmetric} otherwise. The appropriate topological descriptors for $T$ depend on the dimensionality of both $\X$ and $\T$, as well as on whether the field is symmetric. In this work, we focus on 2D tensor fields of the form $T : \R^2 \rightarrow \T$, where $\T$ is the space of $2 \times 2$ matrices. 

\para{Degenerate points.} The \emph{degenerate points} of a symmetric tensor field $T$ are points $x \in \X$ at which the eigenvalues of $T(x)$ coincide. Interpreting the eigenvectors of $T(x)$ as defining a directional field, these points play a role analogous to critical points in vector fields. Degenerate points are typically classified as \emph{trisectors} or \emph{wedges}, depending on the local behavior of the field. We visualize the induced directional field and its degenerate points in \cref{fig:topo-descriptors}(i). See~\cite{delmarcelle1994topology} for a detailed treatment of symmetric tensor field topology. 

For asymmetric tensor fields, degenerate points can be defined analogously as points where the eigenvalues of $T(x) + T(x)^T$ coincide. However, the topology of asymmetric tensor fields is generally not characterized primarily by these points.

\para{Eigenvector and eigenvalue partitions.} For asymmetric tensor fields, we define the \emph{eigenvector partition} and the \emph{eigenvalue partition}~\cite{lin20112d}. These descriptors partition the domain according to how each tensor $T(x)$ acts as a linear operator on $\R^2$. The eigenvector partition classifies each point based on its local rotation behavior and on whether anisotropic stretching or rotational effects dominate; it also incorporates degenerate points in its definition. The eigenvalue partition classifies points into qualitative regimes such as positive scaling, negative scaling, clockwise rotation, counterclockwise rotation, or anisotropic stretching. We visualize these partitions in \cref{fig:tensors}. See~\cite{lin20112d} for a detailed treatment of the topology of asymmetric tensor fields.

\subsection{Feature Tracking}

Topological descriptors often correspond to meaningful structures in data. For time-varying data, it is therefore important to track how these features evolve over time. This is achieved by computing descriptors at each time step and establishing correspondences between features across successive steps. {\tool} supports tracking of critical points for scalar and vector fields, and degenerate points for tensor fields, and provides three strategies for computing these correspondences.

All three strategies are based on optimal transport, which computes optimal matchings between features. In our setting, they determine correspondences between topological features at adjacent time steps. Each method produces a \emph{coupling matrix} representing a soft matching between features, from which we derive a pairing.

{\tool} implements three algorithms from the Python Optimal Transport (POT) library~\cite{flamary2024pot}. The \emph{Earth Mover's Distance} (EMD) computes an exact optimal coupling based on feature distances but is computationally expensive. \emph{Sinkhorn} introduces entropic regularization to approximate EMD more efficiently. The \emph{Partial Wasserstein} method  modifies the objective to better handle differing numbers of features between time steps at the cost of increased computational complexity.

%% file: appendix-additional-results.tex
\section{Experimental Results: time, token, accuracy}

\begin{table}[!ht]
\begin{tabular}{l|ccccc} \hline
model & Time & I.T. & O.T. & B.S. & Acc. \\ \hline
GPT & 20.4 & 27.1K & 657 & 9.9 & 99.2\% \\
GPT (control) & 19.8 & 17.4K & 570 & 8.9 & 50.8\% \\
Claude & 26.6 & 49.9K & 1.4K & 9.9 & 97.5\% \\
Claude (control) & 18.2 & 31.8K & 1.2K & 8.9 & 61.3\% \\
\hline
\end{tabular}
\caption{Average runtime (including rendering time), token usage, and accuracy for {\tool} on fully specified prompts (prompts 1-3) for feasible tasks. GPT (control) and Claude (control) denote
{\tool} using GPT and Claude, respectively, with safeguards disabled.
Time: average total runtime, including computation, in
seconds; I.T./O.T.: average input/output tokens; B.S.: average
benchmark score; Acc.: percentage of trials with no errors.}
\label{tab:additional-experimental-results-fully-specified}
\end{table}

\begin{table}[!ht]
\begin{tabular}{l|ccccc} \hline
model & Time & I.T. & O.T. & B.S. & Acc. \\ \hline
GPT & 23.4 & 33.9K & 753 & 10.0 & 98.4\% \\
GPT (control) & 27.4 & 16.7K & 546 & 9.4 & 47.5\% \\
Claude & 26.0 & 62.4K & 1.5K & 10.0 & 98.8\% \\
Claude (control) & 27.8 & 36.8K & 1.4K & 9.9 & 45.3\% \\
\hline
\end{tabular}
\caption{Average runtime (including rendering time), token usage, and accuracy for {\tool} on moderately specified prompts (prompts 4-7) for feasible tasks. GPT (control) and Claude (control) denote
{\tool} using GPT and Claude, respectively, with safeguards disabled.
Time: average total runtime, including computation, in
seconds; I.T./O.T.: average input/output tokens; B.S.: average
benchmark score; Acc.: percentage of trials with no errors.}
\label{tab:additional-experimental-results-moderately-specified}
\end{table}

\begin{table}[!ht]
\begin{tabular}{l|ccccc} \hline
model & Time & I.T. & O.T. & B.S. & Acc. \\ \hline
GPT & 25.1 & 36.9K & 820 & 10.0 & 99.6\% \\
GPT (control) & 21.1 & 13.8K & 479 & 4.9 & 13.3\% \\
Claude & 26.5 & 64.6K & 1.4K & 10.0 & 99.6\% \\
Claude (control) & 39.3 & 37.6K & 1.5K & 6.3 & 13.8\% \\
\hline
\end{tabular}
\caption{Average runtime (including rendering time), token usage, and accuracy for {\tool} on minimally specified prompts (prompts 8-10) for feasible tasks. GPT (control) and Claude (control) denote
{\tool} using GPT and Claude, respectively, with safeguards disabled.
Time: average total runtime, including computation, in
seconds; I.T./O.T.: average input/output tokens; B.S.: average
benchmark score; Acc.: percentage of trials with no errors.}
\label{tab:additional-experimental-results-minimally-specified}
\end{table}

\begin{table}[!ht]
\begin{tabular}{l|cccc} \hline
model & Time & I.T. & O.T. & Acc. \\ \hline
GPT & 8.3 & 9.4K & 313 & 90.5\% \\
GPT (control) & 11.7 & 12.6K & 397 & 0.0\% \\
Claude & 5.3 & 8.0K & 407 & 100.0\% \\
Claude (control) & 12.6 & 24.5K & 874 & 0.5\% \\
\hline
\end{tabular}
\caption{Average runtime (including rendering time), token usage, and accuracy for {\tool} in infeasible tasks. GPT (control) and Claude (control) denote {\tool} using GPT and Claude, respectively, with safeguards disabled. Time: average total runtime, including computation, in seconds; I.T./O.T.: average input/output tokens.}
\label{tab:additional-experimental-results-infeasible}
\end{table}

\new{In this section, we provide additional experimental details. We report the execution time, token consumption, and accuracy of {\tool} for fully, moderately, and minimally specified prompts in \cref{tab:additional-experimental-results-fully-specified}, \cref{tab:additional-experimental-results-moderately-specified}, and \cref{tab:additional-experimental-results-minimally-specified}, respectively. More specific prompts generally require fewer tokens and less time, as the system needs to gather less information. Across these prompt specificity levels, the accuracy of {\tool} (GPT) remains fairly consistent, while that of {\tool} (Claude) improves slightly. In the control trials, accuracy decreases as prompt specificity decreases, consistent with the need to make more assumptions when less information is provided.}

\new{In \cref{tab:additional-experimental-results-infeasible}, we report the performance of {\tool} on infeasible tasks. Overall, token consumption and execution time are considerably lower than those reported in \cref{tab:agent-results}. When {\tool} correctly identifies a task as infeasible, it generally does so quickly and uses few resources. When it fails to recognize infeasibility, it typically generates relatively simple pipelines, keeping resource consumption low even in the control trials.}

%% file: appendix-additional-case-studies.tex
\section{Additional Case Studies with {\tool}}
\label{appendix:additional-case-studies}

In this section, we list additional case studies using {\tool}.

\para{Extract critical points.}~We use {\tool} to visualize the critical points of the electron density field surrounding a molecule. Topological features often carry meaningful chemical interpretations; for example, high-persistence minima correspond to atomic nuclei. In practice, our collaborating chemists frequently prototype a variety of topology-based visualizations to identify patterns in chemical data, a process that requires making many decisions across multiple stages. {\tool} streamlines this workflow by automatically suggesting relevant options and making it easy to adjust individual components. During the conversation, {\tool} proposes multiple alternatives to the user and uses the user's response to configure a multi-step workflow, thus highlighting its conversational nature. The resulting visualization is shown in \cref{fig:molecule}(b), and the corresponding node tree is displayed in \cref{fig:molecule}(a).

\para{Merge and contour tree.} We use {\tool} to visualize the merge and contour trees of an Ionization dataset, a time-varying scalar field. In this dataset, extrema correspond to areas of high or low velocity. By visualizing the merge and contour trees, we can visualize pockets of the front that correspond to locally fast or slow regions and how they connect geometrically. It is difficult to visualize such features in an interpretable way using standard strategies such as volume rendering. 

\begin{figure}[!ht]
\centering
\includegraphics[width=0.8\linewidth]{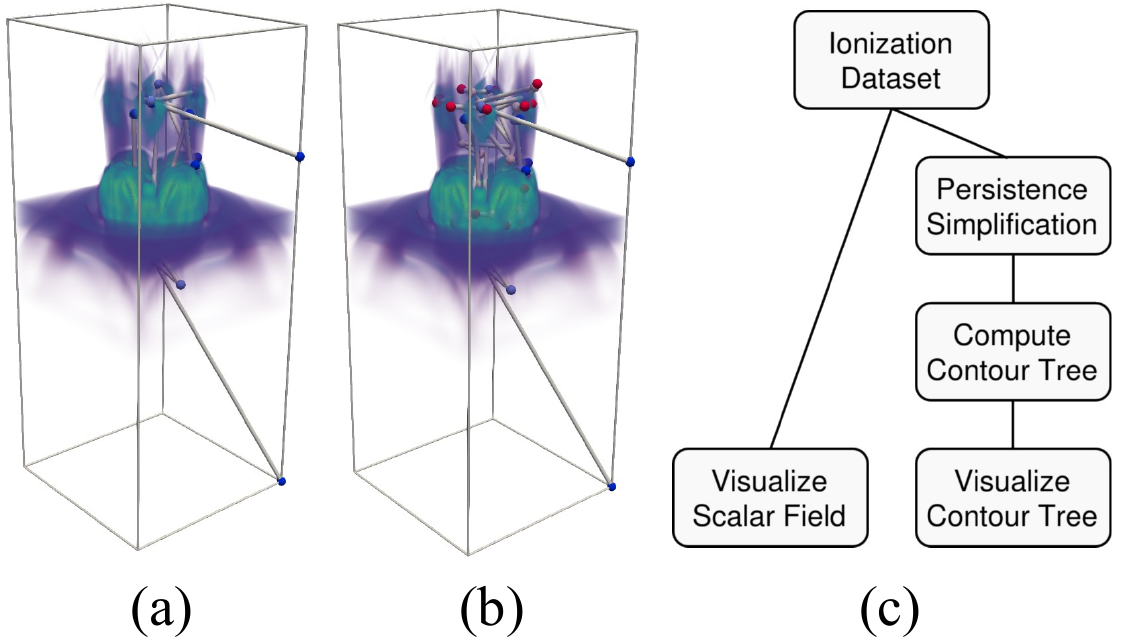}
\vspace{-2mm}
\caption{Computing and simplifying merge trees (a) and contour trees (b) of the Ionization dataset, with (c) node tree diagram for (b).}
\label{fig:atoms-ionization}
\vspace{-4mm}
\end{figure}

During the interaction, the user requests switching from the merge tree to the contour tree. Due to the caching mechanism, the simplified scalar field is not recomputed, illustrating how {\tool} supports rapid prototyping. Moreover, because {\tool} recognizes that the Ionization dataset is time-varying, it handles multiple time steps automatically. The final visualization of the merge tree for the first time step is shown in \cref{fig:atoms-ionization}(a), the contour tree in \cref{fig:atoms-ionization}(b), and the node tree in \cref{fig:atoms-ionization}(c).

%% file: appendix-implementation-details.tex
\section{Implementation Details}
\label{appendix:implementation-details}

In this section, we provide further explanation of {\tool}'s implementation. We describe how the nodes in the node tree are implemented in \cref{appendix:implementation-details-node-implementation}. We describe how the user interface and code generation are performed in \cref{appendix:implementation-details-user-interface}. Finally, we describe our automatic transfer function strategy in \cref{appendix:implementation-details-transfer-function}

\subsection{Node implementation}
\label{appendix:implementation-details-node-implementation}

Each node type is implemented as its own class, inheriting from one of two abstract base classes: \texttt{RootNode} for root node types and \texttt{Node} for all other node types. The implementation of each node type includes the code required to perform its operation, as well as a description of its functionality (including parameter descriptions) that is used by {\tool}. Each node instance contains a five-character ID string that serves as its unique identifier. Constructors for classes inheriting from \texttt{RootNode} take as input a filename and the array names to be loaded. Constructors for classes inheriting from \texttt{Node} take as input the ID of the parent node in the node tree, along with any parameters required to perform the operation (e.g., a persistence threshold).

Because nodes are implemented independently, there are no restrictions on the libraries or frameworks used in their implementation, aside from compatibility with the internal data formats. Several operations are implemented using the Python bindings for ParaView and the Topology Toolkit (TTK)~\cite{tierny2018ttk}. We use additional Python libraries, such as Python Optimal Transport (POT)~\cite{flamary2024pot, flamary2021pot} for feature tracking, as well as a custom C++ backend exposed to Python via pybind11~\cite{jakobPybind}.

During execution, the output of each node is cached to disk. To ensure the correctness of cached results, node parameters (e.g., persistence thresholds) are treated as immutable. An operation may be performed with alternate parameter values by adding a separate node to the node tree. For time-varying datasets, the node tree is evaluated and cached independently for each time step.

\subsection{User Interface and Code Generation}
\label{appendix:implementation-details-user-interface}

\para{User interface.} We implement the visualization interface using Trame VTK \cite{Trame}, a web-based interface for VTK that integrates with ParaView. The chat interface connects to a user-selected foundation model through its API or with the MCP.

\para{Code generation.} {\tool} can deterministically generate code to recreate its visualization. When exporting Python code, we first generate code that reconstructs the node tree. Because each node is implemented as a class, we traverse the tree and generate code that instantiates the corresponding object for each node. This node-tree construction code is then inserted into a boilerplate command-line tool and exported as a standalone script. The resulting script depends only on ParaView (with TTK) and the Python library for {\tool}. The generated node tree is designed to be readable, helping users understand the sequence of computations performed.

\subsection{Automatic Transfer Function Generation}
\label{appendix:implementation-details-transfer-function}

Since transfer function generation is not a central contribution, we adopt a lightweight strategy that we find to be fairly effective. Let $\X$ be a 3D regular grid and $f : \X \rightarrow \R$ the scalar field used for volume rendering. At a high level, our method identifies an interval $I = (a,b)$ such that $\lvert f^{-1}(I) \rvert \approx 0.2\, \lvert \X \rvert$ while maximizing the width $\lvert b - a \rvert$. The opacity transfer function is then centered around this interval. Intuitively, regions of interest tend to correspond to values where the scalar field varies significantly, whereas near-constant regions often represent background or less informative structure. 

Our algorithm computes only the opacity component of the transfer function. The color component is specified by the user via a colormap scaled to the visible scalar range. Pseudocode is provided in \cref{alg:transfer-function}. The resulting transfer function is represented as a list of (function value, opacity) pairs.

\textcolor{black}{
\begin{algorithm}[!ht]
\caption{Automatic transfer function generation. This algorithm computes an opacity transfer function for a scalar field $f : \X \rightarrow \R$ defined on a 3D grid $\X$. The transfer function is returned as a list of pairs of the form (function value, opacity).}
\begin{algorithmic}
\State // sample $\approx \min\left(\left| \X \right|, 10^6\right)$ points
\State $s \gets \max\left( 1, \left\lfloor \frac{ \left| \X \right| }{10^6} \right\rfloor \right)$.
\State $L \gets []$ // empty list
\State $i \gets 0$
\State $\X' \gets$ the points of $\X$ flattened to a 1D array in Fortran order
\While{$i < \left\| \X' \right\|$}
    \State append $\X'[i]$ to $L$
    \State $i \gets i + s$
\EndWhile
\State \quad
\State // find window of points with maximum function difference
\State Sort $L$ by function value.
\State $w \gets \left\lfloor \frac{|L|}{5} \right\rfloor$
\State $b \gets \text{argmax}_x f(L[x+w])-f(L[x])$
\State $m \gets f(L[b])$
\State $r \gets f(L[b+w])-f(L[b])$
\State \quad
\State // construct transfer function stored as a list of
\State  // pairs (function value, opacity)
\State $TF \gets []$ // empty list
\State \quad
\State append $(m,0)$ to $TF$
\State append $(m+0.25r,0.2)$ to $TF$
\State append $(m+0.375r,0.5)$ to $TF$
\State append $(m+0.5r,1.0)$ to $TF$
\State append $(m+0.625r,0.5)$ to $TF$
\State append $(m+0.75r,0.2)$ to $TF$
\State append $(m+r,0)$ to $TF$
\State \Return $TF$
\end{algorithmic}
\label{alg:transfer-function}
\end{algorithm}
}

%% file: appendix-tester.tex
\section{Design of the Tester Agent}

To evaluate {\tool}'s correctness, we use a tester agent to simulate a user. The tester answers follow-up questions from {\tool} and determines when each trial is complete. For each prompt, the tester receives the information it is permitted to provide to {\tool}, represented as key--value pairs. Whenever {\tool} produces a text response, the full conversation history is sent to the tester, which determines whether the orchestrator has requested the value associated with each key. We constrain the tester's output using tool calls so that it returns only a Boolean value for each key, indicating whether that key has been requested. For each requested key, the corresponding value is deterministically returned to {\tool} in the format ``key: value.''

The tester receives only the keys, not their values, preventing it from determining whether {\tool} selects incorrect parameter values. It is also not informed which keys require explicit requests, ensuring that information is provided only when requested. Each key--value pair is returned to {\tool} at most once; if the same key is requested again, no value is returned.

The tester also determines when to terminate a trial. A trial terminates when {\tool} produces a text response and the tester returns no additional key--value pairs. The tester may also terminate a trial early if {\tool} declares the task infeasible. A trial involving an infeasible task is considered correct if {\tool} ultimately concludes that the task is infeasible. If {\tool} initially deems the task feasible but later revises its conclusion, the trial is still counted as correct because it demonstrates self-correction. We manually verify the outcomes of all infeasible-task trials.

For feasible tasks, when {\tool} generates a visualization, we compare the resulting node tree with a ground-truth node tree. A trial is scored as correct if the two node trees imply identical visualizations and as incorrect otherwise. Because we compare the node trees directly, the evaluation does not require executing any computations or rendering the visualizations.

%% file: appendix-evaluation-notes.tex
\section{\new{Evaluation Detail}}

\new{\para{Control trial timeouts.} In some control trials, {\tool} incorrectly configures operations that are computationally expensive, such as large optimal transport problems. To keep experiment durations manageable, we impose a 400-second time limit on each control trial. Of the 2000 control trials, three time out; all others complete in under 180 seconds.}

\para{Interactivity from coding agents.} \new{The coding agents run in headless mode and cannot receive answers to follow-up questions. To support a fair comparison, we allow greater interpretive flexibility when evaluating their responses, whereas only one response is accepted as correct for {\tool}. When evaluated on the same fully specified prompts used for Codex and Claude Code, {\tool} asks no follow-up questions in $99.4\%$ of trials. Thus, in nearly all trials using the same prompts, neither {\tool} nor the coding agents receive additional information.}

\para{SciVisAgentBench evaluation.} We evaluate {\tool} and two widely used coding agents (Codex and Claude Code) using SciVisAgentBench~\cite{ai2026scivisagentbench}. Because the benchmark requires raw data to assess correctness, we modify the generated scripts to export the necessary data. We make these modifications manually for Codex and Claude Code and use an automated tool for {\tool}. In addition, when a pipeline generated by Codex or Claude Code with access to {\tool} is configured correctly---which is generally straightforward to verify---we assign it a score of 10.

SciVisAgentBench has a flag allowing a slight margin of error to score a perfect 10, allowing for algorithms with small differences (e.g., due to rounding) to be considered perfect. We enabled this flag.

We use the following SciVisAgentBench evaluations for each task:

\begin{itemize}[noitemsep,leftmargin=*]

\item \textbf{Critical points and isocontours.} We evaluate the critical points using the benchmark's point-cloud Wasserstein score.

\item \textbf{Contour tree.} Because SciVisAgentBench does not provide a method for evaluating contour trees, we evaluate the tree vertices using the point-cloud Wasserstein score. We perform this evaluation only for the first time step.

\item \textbf{Vector field critical point tracking.} We evaluate the critical points at the first time step using the point-cloud Wasserstein score. We discard critical-point type information because the coding agents are not asked to compute it. SciVisAgentBench does not provide a method for evaluating feature tracking.

\item \textbf{Scalar field critical point tracking.} The coding agents generally compute maxima in a manner closely coupled with the tracking results, making the maxima difficult to export independently. However, SciVisAgentBench cannot evaluate tracking and can evaluate only the detected maxima. Because computing maxima is not the primary objective of this task and critical-point detection is already evaluated in two other tasks, we do not evaluate this task using SciVisAgentBench.

\item \textbf{Morse--Smale segmentation.} We evaluate the Morse--Smale segmentation and its critical points using the corresponding evaluation functions provided by the benchmark.

\item \textbf{Symmetric tensor field.} We evaluate the locations of degenerate points using the benchmark's point-cloud Wasserstein score.

\item \textbf{Asymmetric tensor field.} We evaluate the eigenvector partition and degenerate points using the corresponding evaluation functions provided by the benchmark.

\item \textbf{Persistence diagram.} We evaluate persistence diagrams using a modified version of the Wasserstein-distance metric that the benchmark uses to evaluate persistence pairs in merge trees.

\end{itemize}

\begin{figure*}
\includegraphics[width=\linewidth]{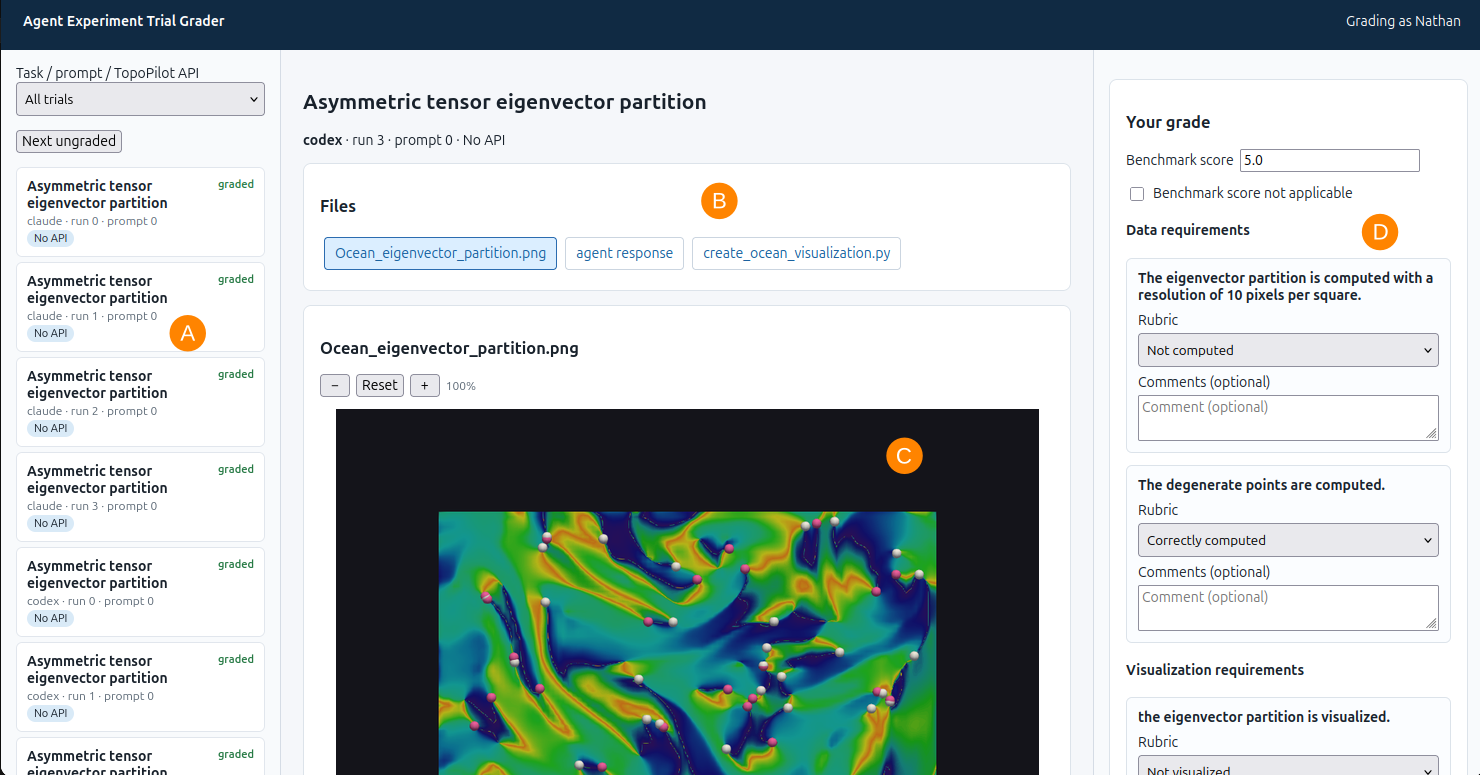}
\caption{Layout of the grading interface. (A) The user selects a trial to grade. (B) The interface lists all files produced by the agent. (C) The display pane renders images or scripts. (D) The rubric enables the user to score each visualization according to how well it satisfies the prompt requirements.}
\label{fig:grader}
\end{figure*}

\para{Manual scoring of coding-agent results.} To score the coding agents' results, we develop a web-based grading application that allows users to inspect each evaluation individually. The application displays every file generated by the agent, including scripts, images, and videos, as well as a file containing the agent's final response to the user.

The user scores each evaluation using a prompt-specific rubric. For each prompt, we identify the computational and visualization requirements and score each requirement separately. For computational requirements, the available ratings are:

\begin{itemize}[noitemsep,leftmargin=*]
\item Correctly computed
\item Correctly computed, but edge cases are not handled correctly
\item Computed with errors
\item Not computed
\end{itemize}

For visualization requirements, the available ratings are:

\begin{itemize}[noitemsep,leftmargin=*]
\item Correctly visualized
\item Visualized, but design choices impair interpretability
\item Visualization requirements are only partially satisfied
\item Not visualized
\end{itemize}

For both types of requirements, the second rating indicates a minor error, whereas the final two ratings indicate major errors. The grader may also provide comments on individual rubric items. \cref{fig:grader} shows a screenshot of the grading application. The application's source code and all scored data are available at 
\url{https://github.com/ngorski9/topopilot-grading-app}.

%% file: appendix-correctness.tex
\section{Details on Guardrails and Mitigation Strategies}
\label{appendix:correctness}

In this section, we enumerate the prompts and guardrails used to ensure the reliability of {\tool}. Node-specific guardrails are listed in \cref{appendix:correctness-guardrails}, the global prompt is provided in \cref{appendix:correctness-global-prompt}, and the verifier prompts are given in \cref{appendix:correctness-verifier}.

\subsection{Node-Specific Guardrails}
\label{appendix:correctness-guardrails}

We list the guardrails that are specific to individual nodes in a node tree:
\begin{itemize}[noitemsep,leftmargin=*]
\item \textbf{Scalar field embedding.} The node that embeds a scalar field is equipped with a guardrail that ensures, when persistence simplification is used, the \emph{unsimplified} scalar field is visualized (rather than the simplified field) unless the user explicitly specifies otherwise.
\item \textbf{Eigenvector partition embedding.} The node that embeds the eigenvector partition of an asymmetric tensor field is equipped with a guardrail that ensures it is always visualized together with the degenerate points.
\item \textbf{Isocontour embedding.} The node that embeds isocontours is equipped with a guardrail that ensures that, when visualized alongside a scalar field, both the contours and the scalar field use the same colormap unless the user explicitly specifies otherwise. We also add a guardrail that ensures that isocontours are computed on the \emph{unsimplified} scalar field, similar to with scalar field embeddings.
\item \textbf{Persistence simplification before computing.} The nodes that compute the Morse--Smale segmentation, scalar field critical points, merge tree, and contour tree are all equipped with guardrails that require explicit user confirmation on whether persistence simplification should be applied to the input. If the user opts for simplification, the guardrail instructs the orchestrator to assume that the computation should use the simplified scalar field as input. 
\item \textbf{Absolute persistence threshold.} We use guardrails to ensure that if the user provides a persistence threshold between 0 and 1, {\tool} assumes that it is relative (rather than absolute) unless they explicitly indicate otherwise.
\item \textbf{Discrete Morse versus piecewise linear.} We use guardrails to ensure that the user wants the piecewise linear version of a topological descriptor, rather than the discrete Morse version, if such a discrepancy exists.
\item \textbf{Limited persistence diagram dimension.} \new{The persistence diagram node uses TTK's FTM backend, which provides persistence diagrams only for dimensions $0$ and $D-1$, where $D$ is the dataset dimension. A guardrail prevents the node from proceeding if the user requests an unsupported diagram dimension.}
\end{itemize}

\subsection{Global Prompt}
\label{appendix:correctness-global-prompt}

During initialization, the global prompt is specified as follows:\\
\begin{quote}
\prompt{
You are an agent designed for performing topological calculations and producing visualizations. The user can only communicate with you by entering text, and cannot upload data or images. You can interact with data on the user's machine by calling tools. Time varying data can be loaded, but it is only viewed one frame at a time (the user has the ability to select which frame they view). You have the following capabilities (as defined in tools):
\\
\begin{itemize}[noitemsep,leftmargin=*]
\item Loading of structured 2D and 3D scalar fields, 2D and 3D vector fields, and 2x2 symmetric and asymmetric 2D structured tensor fields.
\item Visualization of 3D scalar fields with volume rendering
\item Visualization of 2D scalar fields with heatmaps
\item Visualization of 2D vector fields with LIC
\item Visualization of 2D symmetric 2x2 tensor fields with hyperLIC
\item Visualization of 2D asymmetric 2x2 tensor fields with the eigenvector and eigenvalue partition
\item Computation and visualizations of isocontours for scalar fields.
\item Computation and visualization of the Morse-Smale segmentation for 2D scalar fields.
\item Persistence simplification
\item Magnitudes of vector and tensor fields
\item Gradients of scalar and vector fields
\item Computation and visualization of the persistence diagram
\item Computation and visualization of scalar field critical points, vector field critical points, and tensor field degenerate points
\item Tracking scalar field critical points, vector field critical points, and tensor field degenerate points over time
\item Computation and visualization of merge and contour trees
\end{itemize} The current functionality allows you to compute and visualize and export various things, but you will not be able to directly perform analysis on the data outputted by individual tools. To produce a visualization you must use tools to create an embedding and then call visualize\_embedding. If the user does not clarify if they are loading a scalar, vector, or tensor field file, and it is not clear from the context, ask the user to clarify which type they are loading. You only have the abilities that are strictly speaking described with the available tools. If the user requests for something that does not perfectly match the tool description, explicitly clarify what they asked for that is not aligned with any tool descriptions and do not call any tools. Never assume what the user wants if they ask for something ambiguous. Be very conservative about this. If the user does not provide a value for a parameter with a default value, then use the default value.}
\end{quote}

\subsection{Verifier Prompts}
\label{appendix:correctness-verifier}

The verifier is provided with example code representing the proposed pipeline, similar to \cref{fig:code-export-example}, along with descriptions of each parameter and all relevant behavioral checks. It is then instructed to invoke the tool \texttt{verify\_result}, defined as follows:

\begin{quote}
\prompt{Each argument description asks a question. Please answer it by giving an appropriate value to the argument. The current code only shows the most recent visualization, so if the user asks for values relative to a previous iteration, give the generated code the benefit of the doubt. Assume that any logic related to time-varying data is handled automatically unless a function's parameter explicitly says otherwise. If the agent asked about time steps ignore it unless it has to do with a specific parameter whose description asks about time steps. Please be concise with your explanations. If an explanation can be left blank then leave it blank.}
\end{quote}

The following questions are asked in the parameter descriptions:

\prompt{
\begin{itemize}[noitemsep,leftmargin=*]
\item Given the user's prompts, was everything that the user asked for in their most recent request visualized (and nothing else)?
\item For any parameters whose description indicates that their value must be explicitly obtained from the user (not including those with default values) were there any for which the user did not provide a value? If yes, set this to True. If no, set this to False. The agent is allowed to assume values for parameters unless they are explicitly denoted as requiring a value from the user.
\item Are there any parameters which have a default value where the agent used a value other than (a) the default value and (b) a value that the user provided?
\item were all of the lines in the function descriptors that were marked as (guardrail) obeyed? set this to True if yes and False if no.
\end{itemize}
}

%% file: appendix-chat-logs.tex
\section{Complete Conversational Logs for Case Studies}
\label{appendix:chat-logs}

We include the complete logs of user interactions with the agent, using the GPT 5.6 terra low-reasoning model as the backend. The case studies are presented in the order in which they were conducted. As before, user prompts are shown in \myuser{green} and agent messages in \myagent{blue}.

\subsection{Compute and Simplify Persistence Diagrams (\cref{sec:scalar-fields})}

\noindent \myuser{Please load and visualize Isabel.vti with the Kindlmann colormap.}

\myagent{Is Isabel.vti a scalar field? If so, I can load it and render it using the Kindlmann colormap.}

\myuser{It is a scalar field}

\myagent{Loaded Isabel.vti as a scalar field and visualized it with the Kindlmann colormap.}

\myuser{Now please visualize the simplified persistence diagram.}

\myagent{What persistence-simplification threshold should I use? Please provide either:
\begin{itemize}[noitemsep]
\item a relative threshold from 0 to 1 (e.g., 0.05 meaning 5\% of the scalar range), or
\item an absolute persistence threshold, specifying that it is absolute.
\end{itemize}
}

\myuser{Please use a relative threshold of 0.04}

\myagent{Applied piecewise-linear persistence simplification with a relative threshold of 0.04 (4\% of the scalar range), then computed and visualized the resulting persistence diagram.}

\subsection{Extract and Visualize Critical Points (\cref{appendix:additional-case-studies})}

\noindent \myuser{Please load the molecule.vti dataset and visualize it along with its critical points}

\myagent{Is molecule.vti a scalar field dataset? If so, I can load it, compute its scalar critical points, and visualize the volume together with those points.}

\myuser{It is a scalar field}

\myagent{Should I compute all critical points (minima, maxima, and saddles), and should they be computed on the original field or on a persistence-simplified version?}

\myuser{Please show just the minima. Please simplify with a relative threshold of 0.01.}

\myagent{Loaded the scalar field and visualized it with its minima, computed after piecewise-linear persistence simplification using the requested relative threshold of 0.01.}

\subsection{Merge and Contour Tree (\cref{appendix:additional-case-studies})}

\myuser{Please load the Ionization dataset and visualize it along with the merge tree. Use the viridis colormap to visualize the dataset.}

\myagent{Would you like the merge tree computed from the original scalar field, or from a persistence-simplified version? If simplified, please provide the persistence threshold.}

\myuser{Please simplify with a relative threshold of 0.1}

\myagent{Loaded the Ionization scalar field and rendered it with the viridis colormap. The displayed merge tree was computed after piecewise-linear persistence simplification using the requested relative threshold of 0.1.}

\myuser{Now please replace the merge tree with the contour tree}

\myagent{Replaced the merge tree with the full contour tree, computed from the same scalar field simplified at the relative persistence threshold of 0.1. The Ionization volume remains shown with the viridis colormap.}

\subsection{Extract Morse-Smale Segmentation (\cref{sec:scalar-fields})}

\myuser{Please load and visualize fracture.vti}

\myagent{Is fracture.vti a scalar, vector, or tensor field? Please specify the field type so I can load and visualize it appropriately.}

\myuser{It is a scalar field}

\myagent{Loaded and visualized fracture.vti as a 2D scalar field using the default cool–warm colormap.}

\myuser{Now please apply persistence simplification. Visualize the Morse-Smale segmentation along with the critical points.}

\myagent{What persistence threshold should I use for simplification? You may provide either:
\begin{itemize}[noitemsep]
\item a relative fraction of the scalar range (e.g., 0.1), or
\item an absolute persistence value
\end{itemize}
Also, which critical points should be shown: all (minima, maxima, and saddles), minima only, maxima only, or saddles only?}

\myuser{Simplify by 0.05 (relative) and show all critical points}

\myagent{Applied piecewise-linear persistence simplification with a relative threshold of 0.05, then computed and visualized the Morse–Smale segmentation with all minima, maxima, and saddle critical points overlaid.}

\subsection{Scalar Field Critical Point Tracking (\cref{sec:scalar-fields})}

\myuser{Please load the cloud dataset. Compute and track the critical points. Visualize the original scalar field along with the tracked critical points.}

\myagent{Before I compute them, please confirm:
\begin{enumerate}[noitemsep]
\item Which critical points should be included: all critical points (minima, maxima, and saddles), or only a specific type? 
\item Should the scalar field be persistence simplified before computing critical points? If yes, provide the persistence threshold. 
\end{enumerate}
}

\myuser{Show only maxima and apply a relative simplification threshold of 0.5}

\myagent{The scalar field is loaded and persistence-simplified at the requested relative threshold of 0.5.
\\\\
To track the maxima, please specify the tracking backend: emd, sinkhorn, or partial. Also confirm whether to track across all time steps.}

\myuser{Please track 3 time steps and use the partial backend}

\myagent{Visualized the original cloud scalar field with trajectories of persistence-simplified maxima tracked over 3 time steps using the partial backend.}

\myuser{Please use the partial backend}

\myagent{Visualized the original cloud scalar field with trajectories of persistence-simplified maxima tracked over 3 time steps using the partial backend.}

\subsection{Vector Field Visualization (\cref{sec:vector-fields})}

\myuser{Please load dampedOscillator.vti. It is a vector field. Please visualize it along with its critical points.}

\myagent{I need the names of the arrays in dampedOscillator.vti that contain the vector field’s x (u\_array\_name) and y (v\_array\_name) components. Please provide those two array names, and I’ll load the field, compute its critical points, and visualize both together.}

\myuser{The components are given by "u" and "v"}

\myagent{Loaded the vector field and rendered it with its detected critical points overlaid.}

\subsection{Vector Field Critical Point Tracking (\cref{sec:vector-fields})}

\myuser{Please load the data/cylinder dataset. Compute and track the vector critical points. Visualize the vector field along with the tracked critical points.}

\myagent{I can do this, but loading the vector field requires the names of its x- and y-component arrays. Please provide the u\_array\_name and v\_array\_name used in cylinder (or confirm their names if known).}

\myuser{The components are "u" and "v"}

\myagent{The vector field and its critical points have been computed. For tracking, please choose a backend algorithm: emd, sinkhorn, or partial.}

\myuser{Please use the partial backend}

\myagent{How many time steps should be tracked? You can provide a number, or say “all time steps.”}

\myuser{all time steps}

\myagent{Loaded the cylinder vector field, computed its critical points, and tracked them across all time steps using the partial backend. The visualization shows the LIC vector field with the tracked critical points overlaid.}

\subsection{Symmetric Tensor Field Visualization (\cref{sec:tensor-fields})}

\myuser{Please load brain.vti which is a symmetric tensor field. Visualize it along with its degenerate points. Give the degenerate points a radius of 0.8}

\myagent{I need the array names for the tensor components in brain.vti:
\begin{itemize}[noitemsep]
\item A: tensor (1,1) 
\item B: tensor (1,2) / (2,1) 
\item D: tensor (2,2) 
\end{itemize}
Please provide those three array names, and I’ll load the field and visualize it together with its degenerate points at radius 0.8.}

\myuser{The components are "A" "B" and "D"}

\myagent{Loaded the symmetric tensor field from brain.vti using components A, B, and D. The visualization shows the tensor field together with its degenerate points, rendered with radius 0.8.}

\subsection{Asymmetric Tensor Field Visualization (\cref{sec:tensor-fields})}

\myuser{Please load Ocean.vti and visualize its eigenvector partition}

\myagent{To load Ocean.vti as a 2×2 asymmetric tensor field, I need the VTK array names for its four tensor components:
\begin{itemize}[noitemsep]
\item A: (1,1)
\item B: (1,2)
\item C: (2,1)
\item D: (2,2)
\end{itemize}
Please provide those array names, and I’ll load the field and visualize its eigenvector partition.}

\myuser{The array names are "A" "B" "C" and "D"}

\myagent{Loaded Ocean.vti using tensor components A, B, C, and D, and rendered its eigenvector partition. Degenerate points are overlaid: white marks trisectors and pink marks wedges.}

\myuser{Please give the degenerate points a radius of 1}

\myagent{Rendered the degenerate points with radius 1.}

\myuser{Now please visualize the eigenvalue partition instead}

\myagent{Rendered the eigenvalue partition for the Ocean tensor field.}

%% file: appendix-evaluation-prompts.tex
\section{Prompts Used for Evaluation}
\label{appendix:evaluation-prompts}

Here, we list the ten prompts used to evaluate each task, along with the key/value pairs provided to the tester agent. For feasible tasks (tasks 3--10), the first three prompts are fully specified and are also used in the Codex and Claude Code experiments. \new{When evaluating Codex and Claude Code with access to the {\tool} source code, we append the following sentence to each prompt:}

\begin{quote}
\prompt{Please use the TopoPilot2 library (which is installed) to accomplish this task. Instructions are included in TOPOPILOT\_SCRIPTING\_GUIDE.txt}
\end{quote}

\new{When the {\tool} source code is not provided, we append the following sentence instead:}

\begin{quote}
\prompt{The current environment has ParaView, TTK, and Python Optimal Transport installed. Use whatever means necessary to solve the problem and do not ask any followup questions.}
\end{quote}

\new{Without this explicit instruction, the agent frequently fails to complete the task or claims it cannot produce a visualization.}

\subsection*{Task 1: Persistence Barcode Visualization (Infeasible)}

The tester is given the following clarifying information:

\begin{itemize}[noitemsep,leftmargin=*]
\item "color of bars" : "black"
\item "type of field in the file" : "scalar field"
\item "array name of the fields in each file" : "Scalars\_"
\item "dimension of scalar field" : "3D"
\end{itemize}

\noindent We use the following prompts:

\begin{itemize}[noitemsep,leftmargin=*]

\item Please load ./QMCPACK.vti and display its persistence barcode

\item Load ./QMCPACK.vti and visualize the corresponding persistence barcode

\item Please open ./QMCPACK.vti and generate a visualization of its persistence barcode

\item Load the dataset at ./QMCPACK.vti and show its persistence barcode

\item Please read ./QMCPACK.vti and visualize the persistence barcode for the dataset

\item Open ./QMCPACK.vti and display a persistence barcode visualization

\item Please load the VTI file ./QMCPACK.vti and visualize its persistence barcode

\item Import ./QMCPACK.vti and show the resulting persistence barcode

\item Please analyze ./QMCPACK.vti and render its persistence barcode

\item Load ./QMCPACK.vti and create a visualization showing its persistence barcode
\end{itemize}

\subsection*{Task 2: Simplify According to the Gradient (Infeasible)}

The tester is given the following clarifying information:

\begin{itemize}[noitemsep,leftmargin=*]
\item "What I mean by simplify according to the gradient": "Simplify each point where points with a larger gradient get more simplification",
\item "persistence threshold for simplification": "0.04",
\item "data type in the file": "scalar field",
\item "array name": "Scalars\_",
\item "colormap": "viridis"
\end{itemize}

\noindent We use the following prompts:

\begin{itemize}[noitemsep,leftmargin=*]

\item Load ./QMCPack.vti, perform gradient-based simplification, and display the resulting simplified scalar field.

\item Open the file ./QMCPack.vti. Simplify the data according to the gradient and visualize the simplified scalar field.

\item Please read ./QMCPack.vti, apply simplification based on the gradient, and render the simplified scalar field.

\item Import ./QMCPack.vti, execute gradient-guided simplification, and show the simplified scalar field.

\item Load the dataset at ./QMCPack.vti, simplify it based on the gradient, and visualize the resulting scalar field.

\item Access ./QMCPack.vti, apply a gradient-based simplification process, and display the simplified scalar field.

\item Read in ./QMCPack.vti, perform simplification according to gradient values, and visualize the resulting scalar field.

\item Open ./QMCPack.vti and carry out simplification guided by the gradient, then render the simplified scalar field.

\item Load the file ./QMCPack.vti, simplify the scalar data using the gradient as guidance, and visualize the simplified output.

\item Please import ./QMCPack.vti, conduct gradient-driven simplification, and present the simplified scalar field

\end{itemize}

\subsection*{Task 3: Critical Points and Isocontours}

The tester is given the following clarifying information:

\begin{itemize}[noitemsep,leftmargin=*]
\item "data type in file" : "scalar field"
\item "array name" : "Scalars\_"
\item "number of isocontours" : "4"
\item "which field to show contours for" : "original (un-simplified) scalar field",
\item "use simplification for critical points" : "yes"
\item "which critical points to display" : "all critical points"
\item "colormap for critical points" : "default"
\item "radius for critical points" : "default"
\item "colormap for isocontours" : "default"
\item "persistence threshold" : "0.04"
\item "discrete or piecewise linear" : "piecewise linear"
\end{itemize}

\noindent We use the following prompts:

\begin{itemize}[noitemsep,leftmargin=*]

\item Please load ./QMCPACK.vti. Apply persistence simplification using a threshold of 0.04 and visualize all piecewise-linear critical points. Additionally, visualize 4 isocontours dividing the volume into 5 evenly sized regions. Show maxima in red, 2-saddles in orange, 1-saddles in white, and minima in blue.

\item Load ./QMCPACK.vti and perform persistence simplification with a threshold of 0.04. Visualize all of its piecewise-linear critical points along with 4 isocontours dividing the volume into 5 evenly sized regions. Render maxima in red, 2-saddles in orange, 1-saddles in white, and minima in blue.

\item Please open ./QMCPACK.vti, simplify it using a persistence threshold of 0.04, and display all of its piecewise-linear critical points. Also include a visualization of 4 isocontours dividing the volume into 5 evenly sized regions. Display maxima using red, 2-saddles with orange, 1-saddles with white, and minima with blue.

\item Please load ./QMCPACK.vti. Apply persistence simplification and visualize all of its critical points together with its isocontours.

\item Load ./QMCPACK.vti and perform persistence simplification. Visualize the resulting critical points along with the dataset's isocontours.

\item Please open ./QMCPACK.vti, simplify it using persistence simplification, and display all critical points alongside its isocontours.

\item Load the dataset at ./QMCPACK.vti. After applying persistence simplification, visualize all of its critical points together with its isocontours.

\item Please load ./QMCPACK.vti and visualize its isocontours along with its critical points.

\item Load ./QMCPACK.vti and display both its isocontours and critical points.

\item Please open ./QMCPACK.vti and create a visualization showing its isocontours together with its critical points.

\end{itemize}

\subsection*{Task 4: Contour Tree}

The tester is given the following information:

\begin{itemize}[noitemsep,leftmargin=*]
\item "data type in file" : "scalar field"
\item "scalar array name" : "Scalars\_"
\item "apply persistence simplification" : "yes"
\item "persistence threshold epsilon" : "0.1"
\item "sphere radius" : "use default"
\item "sphere colors" : "use default"
\item "edge radius" : "use default"
\item "how to visualize both things" : "at the same time"
\item "colormap for scalar field" : "default"
\item "discrete or piecewise linear" : "piecewise linear"
\item "view contour tree or just the vertices"  : "visualize the full contour tree not just the vertices"
\item "persistence threshold is absolute or relative" : "relative"
\end{itemize}

\noindent We use the following prompts:

\begin{itemize}[noitemsep,leftmargin=*]

\item Load the time-varying Ionization dataset located in ./Ionization (stored as a directory). Visualize the original scalar field with the viridis colormap, then apply persistence simplification with a threshold of 0.1 and display the resulting simplified piecewise-linear contour tree. Draw the maxima in red, 2-saddles in orange, 1-saddles in white, and minima in blue. Give the tree edges a radius of 1 and the vertices a radius of 2.

\item Please import the Ionization time-varying dataset from ./Ionization. First render the original scalar field using the viridis colormap, then perform persistence simplification with value 0.1 and visualize the simplified piecewise-linear contour tree. Give the vertices a radius of 2, rendering the maxima in red, 2-saddles in orange, 1-saddles in white, and minima in blue. Render the edges with a thickness of 1.

\item Open the time-dependent Ionization dataset stored in ./Ionization (directory format). Show the raw scalar field with the viridis colormap, apply persistence simplification with threshold 0.1, and visualize the simplified piecewise-linear contour tree with edges with a radius of 1 and vertices with a radius of 2. Show maxima in red, 2-saddles in orange, 1-saddles in white and minima in blue.

\item Load the dataset located at ../../data/Ionization. Visualize the original scalar field using the viridis colormap, apply persistence simplification, and render the simplified contour tree.

\item Please open ../../data/Ionization, display the original scalar field with the viridis colormap, then perform persistence simplification and visualize the resulting simplified contour tree.

\item Import the dataset from ../../data/Ionization, show the scalar field using the viridis colormap, and apply persistence simplification to generate and visualize the simplified contour tree.

\item Read the dataset at ../../data/Ionization, render the original scalar field using the viridis colormap, and after performing persistence simplification display the simplified contour tree.

\item Load ../../data/Ionization and visualize the dataset with the viridis colormap together with its contour tree.

\item Open the dataset at ../../data/Ionization and display both the scalar field (with the viridis colormap) and its contour tree.

\item Import ../../data/Ionization and visualize the data (using the viridis colormap) along with the corresponding contour tree.

\end{itemize}

\subsection*{Task 5: Vector Field Critical Point Tracking}

The tester is given the following clarifying information:

\begin{itemize}[noitemsep,leftmargin=*]
\item "data type in file" : "vector field"
\item "array names" : "The x component array is 'u' and the y component array is 'v'"
\item "which critical points" : "all critical points"
\item "how many time steps for tracking" : "3"
\item "which backend for tracking" : "partial"
\item "radius for the points" : "default"
\end{itemize}

\noindent We use the following prompts:

\begin{itemize}[noitemsep,leftmargin=*]

\item Load the time-varying cylinder dataset located in ./cylinder (stored as a directory). The piecewise-linear vector field components are in the arrays 'u' and 'v'. Track the critical points over 3 time steps using partial optimal transport and visualize the tracked critical points together with the original vector field.

\item Open the time-dependent cylinder dataset from ./cylinder (directory format). It represents a piecewise-linear vector field with component array names 'u' and 'v'. Use partial optimal transport to track critical points across 3 time steps and display the tracked critical points along with the original vector field.

\item Import the cylinder time-varying dataset in ./cylinder (provided as a directory). The piecewise-linear vector field is defined by component array names 'u' and 'v'. Track its critical points for 3 time steps using partial optimal transport and visualize the tracked points together with the original vector field.

\item Load the dataset from ./cylinder where the x component array name is 'u' and the y component array name is 'v'. Visualize the vector field and track its critical points over time, then display the tracked points.

\item Open ./cylinder, interpreting the array 'u' as the x component and 'v' as the y component of the vector field. Visualize the field and track its critical points over time, rendering the tracked points.

\item Import the dataset located at ./cylinder with vector component arrays 'u' (for x) and 'v' (for y). Visualize the vector field and perform temporal tracking of its critical points, displaying the tracked points.

\item Read the dataset ./cylinder where the array 'u' defines the x component and 'v' defines the y component. Show the vector field and track its critical points through time, visualizing the tracked points.

\item Load ./cylinder and visualize it together with its tracked critical points.

\item Open the dataset at ./cylinder and display it along with the tracked critical points.

\item Import ./cylinder and visualize the data together with the critical points tracked over time.

\end{itemize}

\subsection*{Task 6: Scalar Field Critical Point Tracking}

The tester is given the following clarifying information:

\begin{itemize}[noitemsep,leftmargin=*]
\item "data type in file" : "scalar field"
\item "array name" : "Scalars\_"
\item "which critical points" : "maxima",
\item "use persistence simplification" : "yes"
\item "persistence threshold" : "0.5"
\item "how many time steps for tracking" : "3"
\item "which backend for tracking" : "emd"
\item "radius for the points" : "default"
\item "colormap for scalar field" : "default"
\item "discrete or piecewise linear" : "piecewise linear"
\end{itemize}

\noindent We use the following prompts:

\begin{itemize}[noitemsep,leftmargin=*]
\item Load the time-varying cloud dataset located in ./cloud (stored as a directory). Visualize the original scalar field, then apply persistence simplification with a threshold of 0.5 and track the piecewise-linear maxima across 3 time steps using earth mover's distance. Display the tracked critical points together with the scalar field using a sphere radius of 2 and a warm-cold colormap.

\item Open the time-dependent cloud dataset from ./cloud (directory format). First render the original scalar field, then perform persistence simplification with value 0.5 and track piecewise-linear maxima over 3 time steps using earth mover's distance. Visualize the tracked critical points along with the scalar field using a sphere radius of 2 and a warm-cold colormap.

\item Import the cloud time-varying dataset in ./cloud (provided as a directory). Show the original scalar field, apply persistence simplification with threshold 0.5, and track the piecewise-linear maxima for 3 time steps using earth mover's distance. Visualize the tracked critical points together with the scalar field using a sphere radius of 2 and a warm-cold colormap.

\item Load the dataset from ./cloud, track the maxima over time, and visualize them together with the original scalar field.

\item Open ./cloud, compute and track the maxima through time, and display them alongside the original scalar field.

\item Import the dataset located at ./cloud, track its maxima across time, and visualize the tracked maxima together with the scalar field.

\item Read the dataset from ./cloud, track maxima over time, and render them along with the original scalar field.

\item Load the dataset at ./cloud and visualize it together with its critical points tracked over time.

\item Open ./cloud and display the scalar field along with the critical points tracked through time.

\item Import ./cloud and visualize the data along with its time-tracked critical points.
\end{itemize}

\subsection*{Task 7: Morse--Smale Segmentation}

The tester is given the following clarifying information:

\begin{itemize}[noitemsep,leftmargin=*]
\item "data type in file" : "scalar field"
\item "array name" : "Scalars\_"
\item "which critical points" : "all critical points"
\item "use persistence simplification" : "yes"
\item "persistence threshold" : "0.05"
\item "use persistence simplification for the critical points in addition to the Morse-Smale complex" : "yes, same threshold"
\item "colors for critical points" : "default"
\item "radius for critical points" : "default"
\item "discrete or piecewise linear" : "piecewise linear"
\item "persistence threshold is absolute or relative" : "relative"
\end{itemize}

\noindent We use the following prompts:

\begin{itemize}[noitemsep,leftmargin=*]

\item Load ./fracture.vti, apply a persistence simplification threshold of 0.05, and visualize the piecewise-linear Morse-Smale segmentation of the simplified scalar field together with all critical points, including minima, maxima, and saddles. Give the critical points a radius of 2. Render maxima in red, saddles in white, and minima in blue.

\item Open ./fracture.vti and perform persistence simplification with a value of 0.05. Then display the piecewise-linear Morse-Smale segmentation for the simplified field along with every critical point type: mins, maxes, and saddles. Show maxes in red, saddles in white, and mins in blue with a radius of 2.

\item Read ./fracture.vti, simplify the field using persistence simplification set to 0.05, and render the resulting piecewise-linear Morse-Smale segmentation plus all associated critical points (minimums, maximums, and saddles). Give the critical point markers a radius of 2 and show maximums in red, saddles in white, and minimums in blue.

\item Load ./fracture.vti and display its persistence-simplified Morse-Smale segmentation together with all critical points.

\item Open ./fracture.vti and visualize the Morse-Smale segmentation after persistence simplification, along with the dataset's critical points.

\item Read ./fracture.vti and render its Morse-Smale segmentation on the persistence-simplified field, including the critical points.

\item Import ./fracture.vti and show the persistence simplified Morse-Smale segmentation together with its critical points.

\item Load ./fracture.vti and visualize both its critical points and its Morse-Smale segmentation.

\item Open ./fracture.vti and display the dataset's Morse-Smale segmentation along with all critical points.

\item Read ./fracture.vti and render its critical points together with the corresponding Morse-Smale segmentation.
\end{itemize}

\subsection*{Task 8: Symmetric Tensor Field}

The tester is given the following clarifying information:

\begin{itemize}[noitemsep,leftmargin=*]
\item "data type in file" : "symmetric tensor field",
\item "array names" : "The A B and D components are given by \verb|'A'|, \verb|'B'|, and \verb|'D'|",
\item "colors for degenerate points" : "default",
\item "radius for degenerate points" : "default"
\end{itemize}

\noindent We use the following prompts:

\begin{itemize}[noitemsep,leftmargin=*]
\item Load ./brain.vti, a symmetric piecewise-linear tensor field with component arrays named "A", "B", and "D", and visualize the dataset together with its degenerate points. Give them a radius of 1. Show trisectors in pink and wedges in white.

\item Open ./brain.vti, which represents a symmetric piecewise-linear tensor field with components "A", "B", and "D", and display it along with its degenerate points. Render trisectors in pink and wedges in white all with a radius of 1.

\item Read ./brain.vti (a symmetric piecewise-linear tensor field whose component arrays are "A", "B", and "D") and render it together with all detected degenerate points (with a radius of 1) showing trisectors in pink and wedges in white.

\item Load ./brain.vti, which is a symmetric tensor field, and visualize the dataset along with its degenerate points.

\item Open ./brain.vti and display the symmetric tensor field together with its degenerate points.

\item Read ./brain.vti (a symmetric tensor field) and render it together with the degenerate points in the field.

\item Import ./brain.vti, identified as a symmetric tensor field, and visualize it along with its degenerate points.

\item Load ./brain.vti and visualize the dataset together with its degenerate points.

\item Open ./brain.vti and display it along with all of its degenerate points.

\item Read ./brain.vti and render the dataset together with its degenerate points.
\end{itemize}

\subsection*{Task 9: Asymmetric Tensor Field}

The tester is given the following clarifying information:

\begin{itemize}[noitemsep,leftmargin=*]
\item "data type in file" : "asymmetric tensor field",
\item "array names" : "The A B C and D components are given by \verb|'A'|, \verb|'B'|, \verb|'C'| and \verb|'D'|",
\item "colors for degenerate points" : "default",
\item "radius for degenerate points" : "default",
\item "resolution for eigenvector partition" : "default"
\end{itemize}

\noindent We use the following prompts:

\begin{itemize}[noitemsep,leftmargin=*]

\item Load ./Ocean.vti, an asymmetric piecewise-linear tensor field with component arrays "A", "B", "C", and "D", and visualize the eigenvector partition together with the degenerate points. Give the partition a resolution of 10 pixels per square. Use a radius of 1 for the degenerate points, coloring trisectors in pink and wedges in white.

\item Open ./Ocean.vti, which represents an asymmetric piecewise-linear tensor field whose components are "A", "B", "C", and "D", and display the eigenvector partition (with a resolution of 10 pixels per square) along with its degenerate points. Render trisectors in pink and wedges in white with a radius of 1.

\item Read ./Ocean.vti (an asymmetric piecewise-linear tensor field with component arrays "A", "B", "C", and "D") and render the eigenvector partition (resolution: 10 pixels per square) together with all degenerate points, keeping trisectors pink, wedges white, and giving them a radius of 1.

\item Load ./Ocean.vti, which is an asymmetric tensor field, and visualize its eigenvector partition together with the degenerate points.

\item Open ./Ocean.vti and display the eigenvector partition of the asymmetric tensor field along with its degenerate points.

\item Read ./Ocean.vti (an asymmetric tensor field) and render the eigenvector partition together with the degenerate points.

\item Import ./Ocean.vti, identified as an asymmetric tensor field, and visualize the eigenvector partition with its degenerate points.

\item Load ./Ocean.vti and visualize the eigenvector partition.

\item Open ./Ocean.vti and display its eigenvector partition.

\item Read ./Ocean.vti and render the dataset's eigenvector partition.

\end{itemize}

\subsection*{Task 10: Persistence Diagram}

The tester is given the following clarifying information:

\begin{itemize}[noitemsep,leftmargin=*]
\item "data type in file" : "scalar field",
\item "array name" : "Scalars\_",
\item "apply persistence simplification" : "yes",
\item "persistence threshold" : "0.04",
\item "tube radius" : "default",
\item "ball radius" : "default"
\item "persistence threshold is absolute or relative" : "relative"
\end{itemize}

\noindent We use the following prompts:

\begin{itemize}[noitemsep,leftmargin=*]

\item Load the scalar field located at ./QMCPACK.vti, apply persistence simplification with a threshold of 0.04, and visualize its persistence diagram from the sublevel set filtration showing orders 0 and 2.

\item Open ./QMCPACK.vti as a scalar field, perform persistence simplification with value 0.04, and display the piecewise linear persistence diagram featuring orders 0 and 2 using the sublevel set filtration.

\item Read the scalar field ./QMCPACK.vti, apply persistence simplification set to 0.04, and render the piecewise-linear sub-level set filtration persistence diagram reflecting homology groups of order 0 and 2.

\item Load ./QMCPACK.vti, apply persistence simplification, and visualize the persistence diagram of the simplified scalar field.

\item Open ./QMCPACK.vti and perform persistence simplification, then display the persistence diagram for the resulting simplified scalar field.

\item Read ./QMCPACK.vti, simplify the scalar field using persistence simplification, and render its persistence diagram.

\item Import ./QMCPACK.vti, apply persistence simplification to the scalar field, and visualize the resulting persistence diagram.

\item Load ./QMCPACK.vti and visualize its simplified persistence diagram.

\item Open ./QMCPACK.vti and display the persistence diagram after simplification.

\item Read ./QMCPACK.vti and render the simplified persistence diagram of the scalar field.

\end{itemize}

%% file: refs-topopilot.bib
@book{milnor1963morse,
	author = {Milnor, John Willard},
	publisher = {Princeton university press},
	title = {{Morse} theory},
	year = {1963}}

@article{garyfallidis2014dipy,
	author = {Garyfallidis, Eleftherios and Brett, Matthew and Amirbekian, Bagrat and Rokem, Ariel and Van Der Walt, Stefan and Descoteaux, Maxime and Nimmo-Smith, Ian},
	doi = {10.3389/fninf.2014.00008},
	journal = {Frontiers in Neuroinformatics},
	number = {8},
	title = {{DIPY}, a library for the analysis of diffusion {MRI} data},
	volume = {8},
	year = {2014}}

@article{edelsbrunner2008persistent,
	author = {Herbert Edelsbrunner and John Harer},
	journal = {Contemporary Mathematics},
	pages = {257-282},
	title = {Persistent homology - a survey},
	volume = {453},
	year = {2008}}

@article{tierny2018ttk,
	author = {Tierny, Julien and Favelier, Guillaume and Levine, Joshua A. and Gueunet, Charles and Michaux, Michael},
	doi = {10.1109/TVCG.2017.2743938},
	journal = {IEEE Transactions on Visualization and Computer Graphics},
	number = {1},
	pages = {832-842},
	title = {The {Topology ToolKit}},
	volume = {24},
	year = {2018}}

@article{bilsky2025understanding,
	author = {Bilsky, Joshua and Clark, Aurora E.},
	doi = {10.1063/5.0281156},
	journal = {The Journal of Chemical Physics},
	number = {9},
	title = {Understanding the shape of chemistry data---Applications with persistent homology},
	volume = {163},
	year = {2025}}

@article{hu2021persistent,
	author = {Hu, Yunfeng and Ounkham, Phonemany and Marsalek, Ondrej and Markland, Thomas E. and Krishmoorthy, Bala and Clark, Aurora E.},
	bsdk-url-1 = {https://doi.org/10.3389/fchem.2021.624937},
	doi = {10.3389/fchem.2021.624937},
	journal = {Frontiers in Chemistry},
	title = {Persistent Homology Metrics Reveal Quantum Fluctuations and Reactive Atoms in Path Integral Dynamics},
	volume = {9},
	year = {2021}}

@article{servis2022amphiphile,
	author = {Michael J. Servis and Biswajit Sadhu and L. Soderholm and Aurora E. Clark},
	doi = {10.1016/j.molliq.2021.117743},
	journal = {Journal of Molecular Liquids},
	title = {Amphiphile conformation impacts aggregate morphology and solution structure across multiple lengthscales},
	volume = {345},
	year = {2022}}

@article{carr2003computing,
	author = {Carr, Hamish and Snoeyink, Jack and Axen, Ulrike},
	doi = {10.1016/S0925-7721(02)00093-7},
	journal = {Computational Geometry},
	number = {2},
	pages = {75-94},
	title = {Computing contour trees in all dimensions},
	volume = {24},
	year = {2003}}

@article{tierny2012generalized,
	author = {Tierny, Julien and Pascucci, Valerio},
	doi = {10.1109/TVCG.2012.228},
	journal = {IEEE Transactions on Visualization and Computer Graphics},
	number = {12},
	pages = {2005-2013},
	title = {Generalized topological simplification of scalar fields on surfaces},
	volume = {18},
	year = {2012}}

@article{wu2021ai4vis,
	author = {Wu, Aoyu and Wang, Yun and Shu, Xinhuan and Moritz, Dominik and Cui, Weiwei and Zhang, Haidong and Zhang, Dongmei and Qu, Huamin},
	doi = {10.1109/TVCG.2021.3099002},
	journal = {IEEE Transactions on Visualization and Computer Graphics},
	number = {12},
	pages = {5049-5070},
	publisher = {IEEE},
	title = {{AI4VIS}: Survey on artificial intelligence approaches for data visualization},
	volume = {28},
	year = {2021}}

@article{huang2022flownl,
	author = {Huang, Jieying and Xi, Yang and Hu, Junnan and Tao, Jun},
	doi = {10.1109/TVCG.2022.3209453},
	journal = {IEEE Transactions on Visualization and Computer Graphics},
	number = {1},
	pages = {1200-1210},
	title = {{FlowNL}: Asking the flow data in natural languages},
	volume = {29},
	year = {2023}}

@article{ai2025nli4volvis,
	author = {Ai, Kuangshi and Tang, Kaiyuan and Wang, Chaoli},
	doi = {10.1109/TVCG.2025.3633888},
	journal = {IEEE Transactions on Visualization and Computer Graphics},
	number = {1},
	pages = {46-56},
	title = {{NLI4VolVis}: Natural Language Interaction for Volume Visualization via {LLM} Multi-Agents and Editable {3D} Gaussian Splatting},
	volume = {32},
	year = {2026}}

@inproceedings{peterka2025chatvis,
	author = {Peterka, Tom and Mallick, Tanwi and Yildiz, Orcun and Lenz, David and Quammen, Cory and Geveci, Berk},
	booktitle = {2025 IEEE 15th Symposium on Large Data Analysis and Visualization},
	doi = {10.1109/LDAV68558.2025.00007},
	pages = {22-32},
	title = {{ChatVis}: Large Language Model Agent for Generating Scientific Visualizations},
	year = {2025}}

@article{biswas2025vizgenie,
	author = {Biswas, Ayan and Turton, Terece L. and Ranasinghe, Nishath Rajiv and Jones, Shawn and Love, Bradley and Jones, William and Hagberg, Aric and Shen, Han-Wei and DeBardeleben, Nathan and Lawrence, Earl},
	doi = {10.1109/TVCG.2025.3634655},
	journal = {IEEE Transactions on Visualization and Computer Graphics},
	number = {1},
	pages = {1021-1031},
	title = {{VizGenie}: Toward Self-Refining, Domain-Aware Workflows for Next-Generation Scientific Visualization},
	volume = {32},
	year = {2026}}

@inproceedings{liu2025paraview,
	author = {Liu, Shusen and Miao, Haichao and Bremer, Peer-Timo},
	booktitle = {2025 IEEE Visualization and Visual Analytics},
	doi = {10.1109/VIS60296.2025.00018},
	pages = {61-65},
	title = {{ParaView-MCP}: An Autonomous Visualization Agent with Direct Tool Use},
	year = {2025}}

@article{seo2025automated,
	author = {Seo, Wonduk and Lee, Seungyong and Kang, Daye and An, Hyunjin and Yuan, Zonghao and Lee, Seunghyun},
	doi = {10.48550/arXiv.2502.11140},
	journal = {arXiv preprint arXiv:2502.11140},
	title = {Automated Visualization Code Synthesis via Multi-Path Reasoning and Feedback-Driven Optimization},
	year = {2025}}

@inproceedings{goswami2025plotgen,
	address = {New York, NY, USA},
	author = {Goswami, Kanika and Mathur, Puneet and Rossi, Ryan and Dernoncourt, Franck},
	booktitle = {Companion Proceedings of the ACM on Web Conference},
	doi = {10.1145/3701716.3716888},
	isbn = {9798400713316},
	location = {Sydney NSW, Australia},
	numpages = {5},
	pages = {1672--1676},
	publisher = {Association for Computing Machinery},
	series = {WWW '25},
	title = {{PlotGen}: Multi-Agent {LLM}-based Scientific Data Visualization via Multimodal Retrieval Feedback},
	year = {2025}}

@inproceedings{yang2024matplotagent,
	address = {Bangkok, Thailand},
	author = {Yang, Zhiyu and Zhou, Zihan and Wang, Shuo and Cong, Xin and Han, Xu and Yukun and Liu, Zhenghao and Tan, Zhixing and Liu, Pengyuan and Yu, Dong and Liu, Zhiyuan and Shi, Xiaodong and Sun, Maosong},
	booktitle = {Findings of the Association for Computational Linguistics},
	doi = {10.18653/v1/2024.findings-acl.701},
	editor = {Ku, Lun-Wei and Martins, Andre and Srikumar, Vivek},
	pages = {11789-11804},
	publisher = {Association for Computational Linguistics},
	title = {{MatPlotAgent}: Method and Evaluation for {LLM}-Based Agentic Scientific Data Visualization},
	url = {https://aclanthology.org/2024.findings-acl.701/},
	year = {2024}}

@inproceedings{chen2025coda,
	author = {Zichen Chen and Jiefeng Chen and Sercan O Arik and Misha Sra and Tomas Pfister and Jinsung Yoon},
	booktitle = {The Fourteenth International Conference on Learning Representations},
	title = {{CoDA}: Agentic Systems for Collaborative Data Visualization},
	year = {2026}}

@article{zhao2025proactiveva,
	author = {Zhao, Yuheng and Shu, Xueli and Fan, Liwen and Gao, Lin and Zhang, Yu and Chen, Siming},
	doi = {10.1109/TVCG.2025.3642628},
	journal = {IEEE Transactions on Visualization and Computer Graphics},
	number = {1},
	pages = {451-461},
	title = {{ProactiveVA}: Proactive Visual Analytics with {LLM}-Based {UI} Agent},
	volume = {32},
	year = {2026}}

@article{lange2025generative,
	author = {Lange, Devin and Gao, Shanghua and Sui, Pengwei and Money, Austen and Misner, Priya and Zitnik, Marinka and Gehlenborg, Nils},
	doi = {10.48550/arXiv.2509.16454},
	journal = {arXiv preprint arXiv:2509.16454},
	title = {A Generative {AI} System for Biomedical Data Discovery with Grammar-Based Visualizations},
	year = {2025}}

@article{ferenc2025composable,
	author = {Ferenc Gyarmati, P{\'e}ter and Moritz, Dominik and M{\"o}ller, Torsten and Koesten, Laura},
	doi = {10.48550/arXiv.2509.05721},
	journal = {arXiv preprint arXiv:2509.05721},
	title = {A Composable Agentic System for Automated Visual Data Reporting},
	year = {2025}}

@article{whalen2008ionization,
	author = {Whalen, Daniel and Norman, Michael L},
	doi = {10.1086/524400},
	journal = {The Astrophysical Journal},
	number = {2},
	pages = {664-675},
	publisher = {IOP Publishing},
	title = {Ionization front instabilities in primordial {H II} regions},
	volume = {673},
	year = {2008}}

@article{Haller11,
	author = {Haller, George and Sapsis, Themistoklis},
	doi = {10.1063/1.3579597},
	journal = {Chaos},
	number = {2},
	title = {{Lagrangian} coherent structures and the smallest finite-time {Lyapunov} exponent},
	volume = {21},
	year = {2011}}

@article{gerrisflowsolver,
	author = {S. Popinet},
	journal = {ClusterWorld},
	number = {6},
	title = {Free Computational Fluid Dynamics},
	volume = {2},
	year = {2004}}

@article{vidal2021progressive,
	author = {Vidal, Jules and Guillou, Pierre and Tierny, Julien},
	doi = {10.1109/TVCG.2021.3060500},
	journal = {IEEE Transactions on Visualization and Computer Graphics},
	number = {6},
	pages = {2833-2850},
	title = {A progressive approach to scalar field topology},
	volume = {27},
	year = {2021}}

@article{maack2023parallel,
	author = {Maack, Robin G. C. and Lukasczyk, Jonas and Tierny, Julien and Hagen, Hans and Maciejewski, Ross and Garth, Christoph},
	doi = {10.1109/TVCG.2023.3261981},
	journal = {IEEE Transactions on Visualization and Computer Graphics},
	number = {4},
	pages = {1942-1955},
	title = {Parallel computation of piecewise linear morse-smale segmentations},
	volume = {30},
	year = {2024}}

@article{Guenther17,
	author = {Tobias G{\"u}nther and Markus Gross and Holger Theisel},
	doi = {10.1145/3072959.3073684},
	journal = {ACM Transactions on Graphics},
	location = {Los Angeles, United States},
	number = {4},
	title = {Generic Objective Vortices for Flow Visualization},
	volume = {36},
	year = {2017}}

@article{tian2022comprehensive,
	author = {Tian, Qiyuan and Fan, Qiuyun and Witzel, Thomas and Polackal, Maya N. and Ohringer, Ned A and Ngamsombat, Chanon and Russo, Andrew W and Machado, Natalya and Brewer, Kristina and Wang, Fuyixue and Setsompop, Kawin and Polimeni, Jonathan R. and Keil, Boris and Wald, Lawrence L. and Rosen, Bruce R. and Klawiter, Eric C. and Nummenmaa, Aapo and Huang, Susie Y.},
	doi = {10.1038/s41597-021-01092-6},
	journal = {Scientific Data},
	publisher = {Springer Nature},
	title = {Comprehensive diffusion {MRI} dataset for in vivo human brain microstructure mapping using 300 {mT/m} gradients},
	volume = {9},
	year = {2022}}

@inproceedings{islam2024datanarrative,
	address = {Miami, Florida, USA},
	author = {Islam, Mohammed Saidul and Laskar, Md Tahmid Rahman and Parvez, Md Rizwan and Hoque, Enamul and Joty, Shafiq},
	booktitle = {Proceedings of the 2024 Conference on Empirical Methods in Natural Language Processing},
	doi = {10.18653/v1/2024.emnlp-main.1073},
	editor = {Al-Onaizan, Yaser and Bansal, Mohit and Chen, Yun-Nung},
	pages = {19253-19286},
	publisher = {Association for Computational Linguistics},
	title = {{DataNarrative}: Automated Data-Driven Storytelling with Visualizations and Texts},
	year = {2024}}

@data{oceanData,
	author = {{E.U. Copernicus Marine Service Information}},
	doi = {10.48670/moi-00021},
	title = {Global Ocean Physics Reanalysis},
	year = {2025}}

@dataset{guiltinan2020fractures,
	author = {Guiltinan, Eric. and Santos, Javier E. and Kang, Qinjun and Cardenas, Bayani and Espinoza, Nicolas D.},
	doi = {10.17612/p522-cc94},
	title = {Fractures with variable roughness and wettability},
	year = {2020}}

@inproceedings{gunther2021introduction,
	author = {G{\"u}nther, Tobias and Baeza Rojo, Irene},
	booktitle = {Topological Methods in Data Analysis and Visualization VI},
	doi = {10.1007/978-3-030-83500-2_15},
	pages = {289-326},
	title = {Introduction to vector field topology},
	year = {2021}}

@inproceedings{lin20112d,
	author = {Lin, Zhongzang and Yeh, Harry and Laramee, Robert S and Zhang, Eugene},
	booktitle = {Topological Methods in Data Analysis and Visualization II},
	doi = {10.1007/978-3-642-23175-9_13},
	pages = {191-204},
	title = {{2D} asymmetric tensor field topology},
	year = {2011}}

@inproceedings{li2025tracking,
	author = {Li, Mingzhe and Chatterjee, Dwaipayan and Glassmeier, Franziska and Senf, Fabian and Wang, Bei},
	booktitle = {IEEE Workshop on Topological Data Analysis and Visualization},
	doi = {10.1109/TopoInVis68599.2025.00013},
	pages = {89-99},
	title = {Tracking Low-Level Cloud Systems with Topology},
	year = {2025}}

@inproceedings{delmarcelle1994topology,
	author = {Delmarcelle, Thierry and Hesselink, Lambertus},
	booktitle = {Proceedings Visualization},
	doi = {10.1109/VISUAL.1994.346326},
	organization = {IEEE},
	pages = {140-147},
	title = {The topology of symmetric, second-order tensor fields},
	year = {1994}}

@inproceedings{jeong2024text,
	author = {Jeong, Sangwon and Li, Jixian and Johnson, Chris R and Liu, Shusen and Berger, Matthew},
	booktitle = {IEEE Visualization and Visual Analytics},
	doi = {10.1109/VIS55277.2024.00047},
	pages = {196-200},
	title = {Text-based transfer function design for semantic volume rendering},
	year = {2024}}

@article{liu2024ava,
	author = {Liu, Shusen and Miao, Haichao and Li, Zhimin and Olson, Matthew and Pascucci, Valerio and Bremer, Peer-Timo},
	doi = {https://doi.org/10.1111/cgf.15093},
	journal = {Computer Graphics Forum},
	number = {3},
	pages = {e15093},
	title = {{AVA}: Towards Autonomous Visualization Agents through Visual Perception-Driven Decision-Making},
	url = {https://onlinelibrary.wiley.com/doi/abs/10.1111/cgf.15093},
	volume = {43},
	year = {2024}}

@inproceedings{mallick2024chatvis,
	author = {Mallick, Tanwi and Yildiz, Orcun and Lenz, David and Peterka, Tom},
	booktitle = {Workshops of the International Conference for High Performance Computing, Networking, Storage and Analysis},
	doi = {10.1109/SCW63240.2024.00014},
	pages = {49-55},
	title = {{ChatVis}: Automating Scientific Visualization with a Large Language Model},
	year = {2024}}

@inproceedings{turhan2024digital,
	author = {Turhan, Cinar and Chang, Bernard and Mohamed, Ali and Esteva, Maria and Ketcham, Richard and McClure, James and Prodanovic, Masa},
	booktitle = {International Symposium of the Society of Core Analysts},
	title = {Digital Porous Media Portal for Image Curation, Characterization, Visualization, and Transport Simulation in Porous Media},
	year = {2024}}

@inproceedings{zomorodian2004computing,
	author = {Zomorodian, Afra and Carlsson, Gunnar},
	booktitle = {Proceedings of the twentieth annual symposium on Computational geometry},
	doi = {10.1145/997817.997870},
	pages = {347-356},
	title = {Computing persistent homology},
	year = {2004}}

@inproceedings{zheng2003hyperlic,
	author = {Zheng, Xiaoqiang and Pang, Alex},
	booktitle = {IEEE Visualization 2003.},
	doi = {10.1109/VISUAL.2003.1250379},
	organization = {IEEE},
	pages = {249-256},
	title = {{HyperLIC}},
	year = {2003}}

@misc{scivis2004,
	howpublished = {\url{http://sciviscontest.ieeevis.org/2004/}},
	title = {{The IEEE SciVis Contest}},
	year = {2004}}

@misc{scivis2008,
	howpublished = {\url{http://sciviscontest.ieeevis.org/2008/}},
	title = {{The IEEE SciVis Contest}},
	year = {2008}}

@misc{prodanovic2025digital,
	author = {Prodanovic, M. and Esteva, M. and Ketcham, R. and Chang, B. and Turhan, C. and Gentle, J. and Khan, S. and Belcher, V.},
	doi = {10.17612/FGMN-D889},
	howpublished = {\url{https://digitalporousmedia.org/}},
	title = {{Digital Porous Media Portal (DPMP)} for Publication, Analysis, and Simulation of Porous Media Images},
	year = {2025}}

@misc{flamary2024pot,
	author = {Flamary, R{\'e}mi and Vincent-Cuaz, C{\'e}dric and Courty, Nicolas and Gramfort, Alexandre and Kachaiev, Oleksii and Quang Tran, Huy and David, Laur{\`e}ne and Bonet, Cl{\'e}ment and Cassereau, Nathan and Gnassounou, Th{\'e}o and Tanguy, Eloi and Delon, Julie and Collas, Antoine and Mazelet, Sonia and Chapel, Laetitia and Kerdoncuff, Tanguy and Yu, Xizheng and Feickert, Matthew and Krzakala, Paul and Liu, Tianlin and Fernandes Montesuma, Eduardo},
	title = {{POT} Python Optimal Transport (version 0.9.5)},
	url = {https://github.com/PythonOT/POT},
	year = {2024}}

@article{flamary2021pot,
	author = {R{\'e}mi Flamary and Nicolas Courty and Alexandre Gramfort and Mokhtar Z. Alaya and Aur{\'e}lie Boisbunon and Stanislas Chambon and Laetitia Chapel and Adrien Corenflos and Kilian Fatras and Nemo Fournier and L{\'e}o Gautheron and Nathalie T.H. Gayraud and Hicham Janati and Alain Rakotomamonjy and Ievgen Redko and Antoine Rolet and Antony Schutz and Vivien Seguy and Danica J. Sutherland and Romain Tavenard and Alexander Tong and Titouan Vayer},
	journal = {Journal of Machine Learning Research},
	number = {78},
	pages = {1-8},
	title = {{POT: Python Optimal Transport}},
	url = {http://jmlr.org/papers/v22/20-451.html},
	volume = {22},
	year = {2021}}

@misc{jakobPybind,
	author = {Jakob, Wenzel},
	howpublished = {\url{https://github.com/pybind/pybind11}},
	title = {{Pybind11}},
	year = {2026}}

@article{Trame,
	author = {Jourdain, S. and O'Leary, P. and Schroeder, W.},
	doi = {10.1109/MCG.2025.3540264},
	journal = {IEEE Computer Graphics and Applications},
	month = mar,
	title = {Trame: {{Platform}} Ubiquitous, Scalable Integration Framework for Visual Analytics},
	year = 2025}

@article{ayyamperumal2024current,
	author = {Ayyamperumal, Suriya Ganesh and Ge, Limin},
	doi = {10.48550/arXiv.2406.12934},
	journal = {arXiv preprint arXiv:2406.12934},
	title = {Current State of {LLM} Risks and {AI} Guardrails},
	year = 2024}

@inproceedings{liu2025toolplanner,
	author = {Liu, Yanming and Peng, Xinyue and Cao, Jiannan and Bo, Shi and Zhang, Yuwei and Zhang, Xuhong and Cheng, Sheng and Wang, Xun and Yin, Jianwei and Du, Tianyu},
	booktitle = {The Thirteenth International Conference on Learning Representations},
	title = {Tool-Planner: {{Task}} Planning with Clusters across Multiple Tools},
	year = 2025}

@article{liu2026toolgate,
	archiveprefix = {arXiv},
	author = {Liu, Yanming and Peng, Xinyue and Cao, Jiannan and Wang, Xinyi and Deng, Songhang and Chen, Jintao and Yin, Jianwei and Zhang, Xuhong},
	doi = {10.48550/arXiv.2601.04688},
	eprint = {2601.04688},
	journal = {arXiv preprint arXiv:2601.04688},
	title = {{{ToolGate}}: {{Contract-grounded}} and Verified Tool Execution for Llms},
	year = 2026}

@article{patel2026six,
	author = {Patel, Khush and Surendira, Siva and George, Jithin and Kapale, Shreyas},
	doi = {10.48550/arXiv.2601.22290},
	journal = {arXiv preprint arXiv:2601.22290},
	title = {The {Six Sigma Agent}: Achieving Enterprise-Grade Reliability in {LLM} Systems through Consensus-Driven Decomposed Execution},
	year = 2026}

@article{xian2025reliable,
	archiveprefix = {arXiv},
	author = {Xian, R. Patrick and Gabison, Garry A and Alaa, Ahmed and Riedl, Christoph and Chrysos, Grigorios G.},
	doi = {10.48550/arXiv.2512.07665},
	eprint = {2512.07665},
	journal = {arXiv preprint arXiv:2512.07665},
	title = {Reliable Agent Engineering Should Integrate Machine-Compatible Organizational Principles},
	year = 2025}

@article{balis2026research,
  title={From research question to scientific workflow: Leveraging agentic {AI} for science automation},
  author={Balis, Bartosz and Orzechowski, Michal and Kica, Piotr and Dygas, Michal and Kuszewski, Michal},
  journal={arXiv preprint arXiv:2604.21910},
  year={2026}
}

@article{wang2026planning,
  title={When Planning Fails Despite Correct Execution: On Epistemic Calibration for {LLM}-Based Multi-Agent Systems},
  author={Wang, Zehao and Jin, Shilong and Cao, Zhao and Wang, Lanjun},
  journal={arXiv preprint arXiv:2605.23414},
  year={2026}
}

@article{liu2026scifi,
  title={{SciFi}: A Safe, Lightweight, User-Friendly, and Fully Autonomous Agentic {AI} Workflow for Scientific Applications},
  author={Liu, Qibin and Gonski, Julia},
  journal={arXiv preprint arXiv:2604.13180},
  year={2026}
}

@misc{zou2026amtfv,
      title={{AMTFV}: Agentic Mathematical Tool-Flow Verification for {LLM} Self-Correction}, 
      author={Rui Zou and Yutao Zhu and Mengqi Wei and Ji-Rong Wen},
      year={2026},
      eprint={2607.29549},
      archivePrefix={arXiv},
      primaryClass={cs.AI},
      url={https://arxiv.org/abs/2607.29549}, 
}

@article{ai2026scivisagentbench,
  title={{SciVisAgentBench}: A benchmark for evaluating scientific data analysis and visualization agents},
  author={Ai, Kuangshi and Miao, Haichao and Tang, Kaiyuan and Gorski, Nathaniel and Sun, Jianxin and Liu, Guoxi and Ingolfsson, Helgi I and Lenz, David and Guo, Hanqi and Yu, Hongfeng and others},
  journal={arXiv preprint arXiv:2603.29139},
  year={2026}
}

@inproceedings{zhao2020sdrbench,
    address = {New York},
	author = {Zhao, Kai and Di, Sheng and Lian, Xin and Li, Sihuan and Tao, Dingwen and Bessac, Julie and Chen, Zizhong and Cappello, Franck},
	booktitle = {2020 IEEE International Conference on Big Data (Big Data)},
	doi = {10.1109/BigData50022.2020.9378449},
	pages = {2716-2724},
    publisher = {IEEE},
	title = {{SDRBench}: Scientific data reduction benchmark for lossy compressors},
	year = {2020}}

@article{kim2018qmcpack,
	author = {Jeongnim Kim and Andrew D. Baczewski and Todd D. Beaudet and Anouar Benali and M. Chandler Bennett and Mark A. Berrill and Nick S. Blunt and Edgar Josu{\'e} Landinez Borda and Michele Casula and David M. Ceperley and Simone Chiesa and Bryan K Clark and Raymond C. Clay III and Kris T. Delaney and Mark Dewing and Kenneth P. Esler and Hongxia Hao and Olle Heinonen and Paul R. C. Kent and Jaron T. Krogel and Ilkka Kyl{\"a}np{\"a}{\"a} and Ying Wai Li and M. Graham Lopez and Ye Luo and Fionn D. Malone and Richard M. Martin and Amrita Mathuriya and Jeremy {McMinis} and Cody A. Melton and Lubos Mitas and Miguel A. Morales and Eric Neuscamman and William D. Parker and Sergio D. Pineda Flores and Nichols A. Romero and Brenda M. Rubenstein and Jacqueline A. R. Shea and Hyeondeok Shin and Luke Shulenburger and Andreas F. Tillack and Joshua P. Townsend and Norm M. Tubman and Brett Van Der Goetz and Jordan E. Vincent and D. {ChangMo} Yang and Yubo Yang and Shuai Zhang and Luning Zhao},
	doi = {10.1088/1361-648X/aab9c3},
	journal = {Journal of Physics: Condensed Matter},
	number = {19},
    pages = {195901:1--195901:29},
	title = {{QMCPACK}: an open source \emph{ab initio} quantum {Monte Carlo} package for the electronic structure of atoms, molecules and solids},
	volume = {30},
	year = {2018}}

@article{kent2020qmcpack,
	author = {Paul. R. C. Kent and Abdulgani Annaberdiyev and Anouar Benali and M. Chandler Bennett and Edgar Josu{\'e} Landinez Borda and Peter Doak and Hongxia Hao and Kenneth D. Jordan and Jaron T. Krogel and Ilkka Kyl{\"a}np{\"a}{\"a} and Joonho Lee and Ye Luo and Fionn D. Malone and Cody A. Melton and Lubos Mitas and Miguel A. Morales and Eric Neuscamman and Fernando A. Reboredo and Brenda Rubenstein and Kayahan Saritas and Shiv Upadhyay and Guangming Wang and Shuai Zhang and Luning Zhao},
	doi = {10.1063/5.0004860},
	journal = {The Journal of Chemical Physics},
	number = {17},
    pages = {174105:1--174105:20},
	title = {{QMCPACK}: Advances in the development, efficiency, and application of auxiliary field and real-space variational and diffusion quantum {Monte Carlo}},
	volume = {157},
	year = {2020}}
